\documentclass[twocolumn]{aastex701}
\usepackage{orcidlink}

\usepackage{amsmath}
\usepackage[version=4]{mhchem} 
\usepackage{placeins}

\allowdisplaybreaks     

\newcommand{\revisedtext}[1]{\textcolor{black}{#1}}

\begin{document}

\title{A New Global Chemical Equilibrium Code: Refractory Element Signatures in Super-Earths and Sub-Neptunes}

\correspondingauthor{Caroline Dorn}

\author[orcid=0000-0002-0632-4407, sname='Grimm']{Simon L. Grimm}
\altaffiliation{These authors contributed equally to this study}
\affiliation{Institute for Particle Physics and Astrophysics, ETH Zurich, CH-8093 Zurich, Switzerland}
\affiliation{Department of Astrophysics, University of Zurich, CH-8057 Zurich, Switzerland}
\email{sigrimm@ethz.ch}  

\author[orcid=0000-0003-0605-0263, sname='Steinmeyer']{Marie-Luise Steinmeyer}
\altaffiliation{These authors contributed equally to this study}
\affiliation{Institute for Particle Physics and Astrophysics, ETH Zurich, CH-8093 Zurich, Switzerland}
\email{}

\author[orcid=0009-0005-1133-7586, sname='Werlen']{Aaron Werlen}
\altaffiliation{These authors contributed equally to this study}
\affiliation{Institute for Particle Physics and Astrophysics, ETH Zurich, CH-8093 Zurich, Switzerland}
\affiliation{Department of Earth, Planetary, and Space Sciences, University of California, Los Angeles, CA 90095, USA}
\email{werlen@ucla.edu}  

\author[orcid=0000-0001-6110-4610, sname='Dorn']{Caroline Dorn}
\affiliation{Institute for Particle Physics and Astrophysics, ETH Zurich, CH-8093 Zurich, Switzerland}
\email[show]{dornc@ethz.ch}

\author[orcid=0000-0002-0298-8089, sname='Schlichting']{Hilke E. Schlichting}
\affiliation{Department of Earth, Planetary, and Space Sciences, University of California, Los Angeles, CA 90095, USA}
\email{hilke@ucla.edu}

\author[orcid=0000-0002-1299-0801, sname='Young']{Edward D. Young}
\affiliation{Department of Earth, Planetary, and Space Sciences, University of California, Los Angeles, CA 90095, USA}
\email{eyoung@epss.ucla.edu}

\begin{abstract}

The atmospheres of super-Earths and sub-Neptunes can be strongly modified by chemical exchange with their molten interiors during long-lived magma ocean phases. Interpreting atmospheric observations requires fast models that self-consistently couple atmospheric chemistry to the composition of the planetary interior. We present an updated, \revisedtext{CPU-based} implementation of the global chemical equilibrium  (GCE) framework from \citet{schlichting_chemical_2022}, which computes the equilibrium composition of a coupled metal–silicate–gas system. The numerical solver has been improved using a gradient-based optimizer, reducing the computational cost of solving the chemical network by more than two orders of magnitude and enabling large parameter studies. We apply the framework to a large synthetic population of planets and explore the imprint of bulk refractory composition of Mg, Si, and Fe on atmospheric properties \revisedtext{in the end member case that hydrogen dissolution increases unbound with pressure}. We consider planets with different masses, thermal states, and volatile inventories. We find that the atmospheric mass fraction and atmospheric metal mass fraction are primarily controlled by the temperature at the atmosphere–magma ocean interface and the planetary water budget, while the accreted hydrogen mass fraction plays a minor role because most hydrogen dissolves into the interior. For planets that accreted water, the refractory ratios Mg/Si and Fe/Si strongly influence carbon partitioning between the gas, silicate, and metal phases, producing large variations in atmospheric metal mass fraction and C/O ratios. These results demonstrate that atmospheric compositions of sub-Neptunes depend sensitively on both the volatile inventory and the bulk composition of rocky material, providing new constraints for interpreting atmospheric observations. The new GCE code is open-source.
\end{abstract}

\linenumbers
\section{Introduction}

In recent years, the chemical composition of super-Earths and sub-Neptunes has become a central topic in exoplanet research. This progress has been driven by major advances in observational capabilities. In particular, observations with the \textit{James Webb Space Telescope} (JWST) have revealed carbon-, oxygen-, and sulfur-bearing species in the atmospheres of several sub-Neptunes \citep[e.g.,][]{madhusudhan_carbon-bearing_2023,benneke_jwst_2024,schmidt_comprehensive_2025,felix_competing_2025,hu_k2-18b_2025,Davenport_toi421b_2025}. At the same time, new theoretical models have been developed to link atmospheric compositions to the interior structure and formation history of these planets \citep[e.g.,][]{schlichting_chemical_2022,seo_role_2024,lichtenberg_constraining_2025,bower_diversity_2025}. Together, these developments are beginning to provide a more detailed picture of the physical and chemical processes that shape the atmospheres of sub-Neptunes.

Unlike smaller terrestrial planets, super-Earths and sub-Neptunes that accrete a few weight (wt)\% of hydrogen and helium are expected to host long-lived magma oceans. The insulating effect of a massive atmosphere can strongly delay cooling of the underlying magma ocean \citep[e.g.,][]{lopez_understanding_2014,ginzburg_super-earth_2016}, allowing molten interiors to persist for gigayear timescales \citep{misener_atmospheres_2023,calder_most_2025}. During these prolonged magma ocean phases, compositional coupling and chemical exchange between the atmosphere and the planetary interior can substantially modify atmospheric composition \citep[e.g.,][]{kite_superabundance_2019,kite_atmosphere_2020,lichtenberg_redox_2021,dorn_hidden_2021,schlichting_chemical_2022,young_differentiation_2025,misener_atmospheres_2023,seo_role_2024,luo_interior_2024,lichtenberg_constraining_2025,werlen_atmospheric_2025,lee_mineral_2025,werlen_sub-neptunes_2025,steinmeyer_coupled_2026}. Understanding these interactions is therefore essential for interpreting atmospheric observations of sub-Neptunes.

One approach to model these processes is the global chemical equilibrium framework introduced by \citet{schlichting_chemical_2022}, which treats the atmosphere, silicate melt, and metal phase as a chemically coupled system. In this framework, the equilibrium composition of all phases is obtained by simultaneously solving a set of equilibrium and mass balance equations. While this approach provides a physically consistent description of interior--atmosphere interactions, the original implementation relied on Markov Chain Monte Carlo (MCMC) techniques, which limited the computational efficiency of the model.


Recent advances in numerical optimization methods allow these equations to be solved more efficiently \citep{kingma+2017}. In this work, we present an updated implementation of the global chemical equilibrium framework that employs a new numerical optimizer, reducing the computational cost of solving the chemical network by more than two orders of magnitude. This improvement enables large parameter studies that were previously computationally prohibitive. The same implementation has already been applied in recent studies of interior--atmosphere interactions in sub-Neptunes \citep{werlen_atmospheric_2025,werlen_sub-neptunes_2025,werlen_effects_2026, steinmeyer_coupled_2026}.

Using this framework, we explore how the atmospheric composition of super-Earths and sub-Neptunes depends on their bulk composition and thermal state. In particular, we investigate how the atmospheric metal mass fraction varies for planets formed inside and outside the ice line as a function of the refractory Mg/Si and Fe/Si ratios. Although these refractory element ratios cannot be measured directly for exoplanets, it has been proposed that they can be informed from the corresponding stellar photospheric abundances \citep{dorn_can_2015}. However, it remains debated whether stellar ratios are reliable proxies for planetary compositions, and whether any correlation is strictly one-to-one \citep{plotnykov_chemical_2020,adibekyan_compositional_2021,schulze_probability_2021,brinkman_revisiting_2024}. Overall, planetary refractory element ratios are likely inherited from their host stars to first order, but the strength and fidelity of this link remain uncertain. Here, we vary bulk Fe/Si and Mg/Si ratios of planets in ranges that may be directly informed from stellar proxies.

The paper is structured as follows. Section~\ref{sec_chemicalmodel} describes the thermodynamic formulation of the global chemical equilibrium framework, including the governing equilibrium and mass balance equations, the adopted reaction network, and the numerical implementation of the solver. Section~\ref{sec_results} presents the results of applying this framework to a synthetic population of super-Earth and sub-Neptune planets. We explore how the atmospheric mass fraction, atmospheric metal mass fraction, and C/O ratio depend on the thermal state of the planet, the accreted volatile inventory, and the refractory composition of the rocky material. Section~\ref{sec_discussion} discusses the physical assumptions and limitations of the framework, including potential deviations from global equilibrium due to interior dynamics or atmospheric structure, as well as uncertainties in the available thermodynamic data \revisedtext{and the influence of the treatment of hydrogen solubility}. Finally, Section~\ref{sec_conclusion} summarizes the main implications of our results for interpreting atmospheric observations of super-Earths and sub-Neptunes.





\section{The Global Chemical Equilibrium Framework}
\label{sec_chemicalmodel}

We adopt the planetary chemical equilibrium (GEC) framework introduced by \citet{schlichting_chemical_2022, Young+2023}, which describes global chemical equilibrium between a molten interior and an atmosphere. The interior consists of a silicate and a metal phase that equilibrate with a gaseous phase at the atmosphere--magma ocean interface (AMOI), characterized by the AMOI temperature. Within the interior, chemical exchange between the silicate and metal phases is governed by the silicate--metal equilibrium (SME) temperature, defined as the temperature at which silicate and metal are assumed to chemically equilibrate. Chemical equilibrium is enforced through a set of linearly independent reactions spanning the relevant reaction space.

The full set of governing equations, as well as a general procedure for constructing the reaction network, is described in \citet{schlichting_chemical_2022}. Here we summarize the equilibrium and mass balance equations relevant for this study.

For global equilibrium, the equilibrium condition must be satisfied for each independent reaction:
\begin{equation}
\label{eq_equilibrium}
\sum_i \nu_i \ln x_i
+ \left[
\frac{\Delta \hat G_r^\circ}{RT}
+ \sum_g \nu_g \ln \left( \frac{P}{P^\circ} \right)
\right]
= 0,
\end{equation}
where $x_i$ is the mole fraction of species $i$, $\nu_i$ its stoichiometric coefficient, $\Delta \hat G_r^\circ$ the standard-state Gibbs free energy of reaction, $R$ the universal gas constant, $T$ the temperature, \revisedtext{$P$ is the pressure at the AMOI, and $P^\circ$ is the standard-state pressure (set to 1 bar)}. The second summation runs over gaseous species (indexed by $g$), for which the explicit pressure dependence is included.

The pressure $P$ is calculated self-consistently during the optimization \revisedtext{following \citet{young_earth_2023}} and is given by
\begin{equation}
\label{eq_pressure}
\frac{P}{1\,\mathrm{bar}}
=
1.2 \times 10^6
\frac{M_{\mathrm{atm}}}{M_{\rm \color{blue}{c}}}
\left(
\frac{M_{\rm c}}{M_\oplus}
\right)^{2/3},
\end{equation}
where $M_{\mathrm{atm}}$ is the atmospheric (gas) mass and $M_{\rm c}$ equals $M_{\rm  pl} - M_{\rm atm}$, all expressed in Earth masses $M_\oplus$.
\revisedtext{This expression assumes that the condensed, non-atmospheric part of the planet sets the planetary radius and surface gravity, and that gravity is approximately constant over the atmosphere. It should therefore be interpreted as an approximate pressure estimate at the atmosphere--magma ocean interface rather than as a full atmospheric structure model. The approximation becomes less accurate for large atmospheric mass fractions, for which the gas envelope may contribute significantly to the planet radius and gravity profile.}

In addition to the equilibrium conditions, elemental mass balance is enforced for each element $s$:
\begin{equation}
\label{eq_mass_balance}
1
-
\sum_k \sum_i
\frac{\eta_{s,i,k} \, x_{i,k} \, N_k}{n_s}
=
0,
\end{equation}
where $n_s$ is the total number of moles of element $s$, $\eta_{s,i,k}$ is the number of atoms of element $s$ in species $i$ of phase $k$, $x_{i,k}$ is the mole fraction of species $i$ in phase $k$, and $N_k$ is the total number of moles in phase $k$.

Finally, a normalization constraint is imposed for each phase:
\begin{equation}
\label{eq_mole_fraction}
\sum_i x_{i,k} = 1.
\end{equation}

The system is solved simultaneously for all mole fractions $x_{i,k}$ and phase mole numbers $N_k$. In contrast to \citet{schlichting_chemical_2022}, Equation~\ref{eq_mass_balance} is normalized by $n_s$, which balances its contribution relative to the equilibrium conditions. \revisedtext{In addition, although the pressure is computed self-consistently in both implementations, we do not treat it as an independent optimization variable. Instead, at each optimization step, $P$ is calculated directly from the current atmospheric and condensed masses using Equation~\ref{eq_pressure}. This reduces the dimensionality of the nonlinear system by one variable compared to the original MCMC implementation.} The resulting nonlinear system of equations is solved using the ADAM optimizer \citep{kingma+2017}. Details of the reaction network, thermodynamic data, and numerical implementation are provided in Appendix~\ref{ap:gce}. Convergence is evaluated using the squared sum of the residuals of the coupled non-linear system of equations (Appendix~\ref{ap:gce}). Here, we consider solutions to be converged when the residual metric falls below $10^{-4}$, although most solutions go well below $10^{-7}$.

A key parameter in our chemical equilibrium calculations is the hydrogen solubility in the silicate melt, as it strongly affects the partitioning of hydrogen between the interior and the atmosphere, and thus the predicted atmospheric mass fractions and compositions, especially the formation of \ce{H2O} and \ce{CH4} in the gas phase. 
\revisedtext{In this work, we employ a fixed equilibrium treatment, which allows the dissolved \ce{H2} abundance to increase unbound with pressure. 
Therefore, the main results presented here should be interpreted within this framework. 
We compare our results to the fixed-$k_\mathrm{D}$ treatment employed in \citet{schlichting_chemical_2022} in Section~\ref{ssec_influence} and Appendix~\ref{app_comp}. The} implementation and thermodynamic basis \revisedtext{of the two different treatments} are described in detail in Appendix~\ref{ap:data}.

The network configuration adopted here is identical to that used in \citet{werlen_atmospheric_2025, werlen_sub-neptunes_2025} and extends the original formulation of \citet{schlichting_chemical_2022} by including carbon partitioning into the metal phase.

The GCE code is open source and publicly available on GitHub\footnote{\url{https://github.com/ExoInteriors/GlobalChemicalEquilibrium_Release}} with full documentation and example workflows. The code is modular and optimized for CPU execution, allowing alternative reaction sets or additional species to be incorporated without modification of the core solver.

In the following sections, using this new code, we explore how the bulk composition and thermal state of a planet influence the atmospheric mass fraction (Section \ref{sec:Matm}) and composition (Section \ref{sec:Zatm}, \ref{sec:abundance}, and \ref{sec:CO}).

\subsection{Example usage: Synthetic planet population}
Our synthetic planet population consists of planets with masses between $1\,M_\oplus$ to $10\,M_\oplus$.
\revisedtext{In order to set the bulk compositions, we assume here that the planets are composed of three primary building blocks, }rocks, water, and primordial gas \revisedtext{with varying mass fractions}. 
\revisedtext{Following \citet{werlen_atmospheric_2025}, we use chondritic abundances of C, O, Na, and Si for the rock composition.}
In order to investigate the effect of rock composition on the atmosphere of a planet, we vary the absolute molar ratios of Mg to Si and Fe to Si in the considered rock material from 0.5 to 2 \revisedtext{using the observed stellar variability to define the boundaries of our parameter space} \citep{hinkel_stellar_2014}. \revisedtext{For reference, the solar molar ratios are (Mg/Si)$_\odot$=1.1 and (Fe/Si)$_\odot$=0.89 \citep{2021AsplundSun}}.
The primordial gas is composed of $99.9\,wt\%$ H$_2$ and $0.1\,wt\%$ CO$_2$ by mole, corresponding to a solar C/O ratio \citep{2018Suarez-AndresCOMg}. 
\revisedtext{Because the contribution of the primordial gas component to the total abundance of elements other than hydrogen is minor, }we refer to the mass fraction of primordial gas as the mass fraction of accreted \ce{H2}, $w_{\mathrm{H}_2}$, and vary it from $0.5\,wt\%$ to $10\,wt\%$ of the total planet mass.
\revisedtext{Consequently, varying the composition of the non-hydrogen component in the primordial gas, for example by using a mixture of CO and \ce{H2O} as has been done in \citet{nixon_magma_2025}, will not influence the results presented in this work. 
We further consider four different water mass fractions, $w_{\mathrm{H}_2\mathrm{O}}=0.0 wt\%, 10wt\%, 20wt\%$, and 50wt\%. 
The effect of incorporating nitrogen and sulfur in the global chemical equilibrium are explored in \citet{2026WerlenFormation} and are not a focus of this study.}
\revisedtext{We assume that the planets are able to reach a state of global chemical equilibrium, which is not dependent on any initial partition of elements.}

For each composition configuration, we calculate the global chemical equilibrium for two different temperatures at the AMOI, $T_\mathrm{AMOI}=2000\,$K and $3000\,$K and three different temperature differences between the AMOI and SME, $\Delta T = 50\,$K, 500\,K, and 1000\,K. 
\revisedtext{The SME temperature indicates how deep in the magma ocean the silicate-metal equilibration takes place. 
As the range of depths (and temperatures) is not well constrained, we treat $\Delta T$ as a free parameter.}
\section{Results}
\label{sec_results}
\label{sec_results}
\begin{figure}
    \centering
    \includegraphics[width=1.0\linewidth]{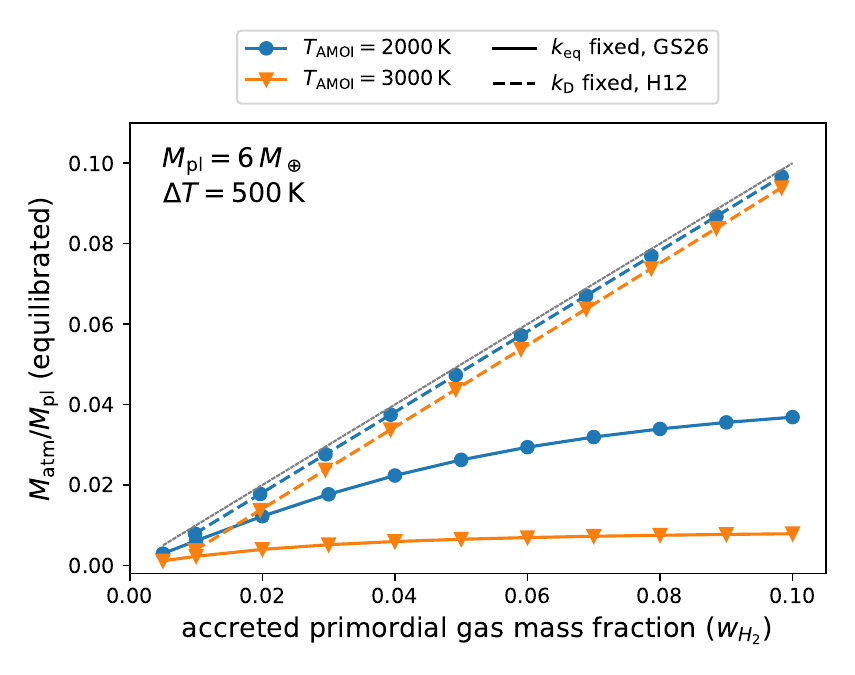}
    \caption{\revisedtext{Atmospheric mass fraction as a function of the accreted mass fraction of primordial gas ($w_{\ce{H2}}$) for two different AMOI temperatures. For all planets, $M_\mathrm{pl}=6\,M_\oplus$, $\Delta T=500\,$K, and $w_{\ce{H2O}}=0.0wt\%$. }}
    \label{fig:Matm_H2_TAMOI_DeltaT}
\end{figure}
\subsection{Atmospheric mass fractions}
\label{sec:Matm}
Figure~\ref{fig:Matm_H2_TAMOI_DeltaT} shows the atmospheric mass fraction as a function of the total accreted \ce{H2} mass fraction ($w_{\ce{H2}}$) and the temperature at the atmosphere--magma ocean interface (AMOI) for a planet with $M_\mathrm{pl}=6\,M_\oplus$ and $w_{\ce{H2O}}=0.0wt\%$.
The temperature difference between the AMOI and the silicate--metal equilibration temperature has only a weak influence on the atmospheric mass fraction and is therefore fixed to $\Delta T=500\,$K.

Hydrogen solubility scales with the AMOI temperature \citep{werlen_effects_2026}, resulting in a strong decrease in the atmospheric mass fraction with increasing $T_\mathrm{AMOI}$. 
For hot or young planets with $T_\mathrm{AMOI} = 3000\,$K, the atmospheric mass fraction $M_\mathrm{atm}/M_\mathrm{pl}$ is reduced to below $\approx 1\,\%$; independent of the accreted \ce{H2} mass fraction. As sub-Neptunes cool and $T_\mathrm{AMOI}$ decrease, atmospheric mass fractions may thus build up over time due to exsolution of volatiles as suggested by \citep{steinmeyer_coupled_2026}.

\begin{figure*}
    \centering
    \includegraphics[width=1.0\linewidth]{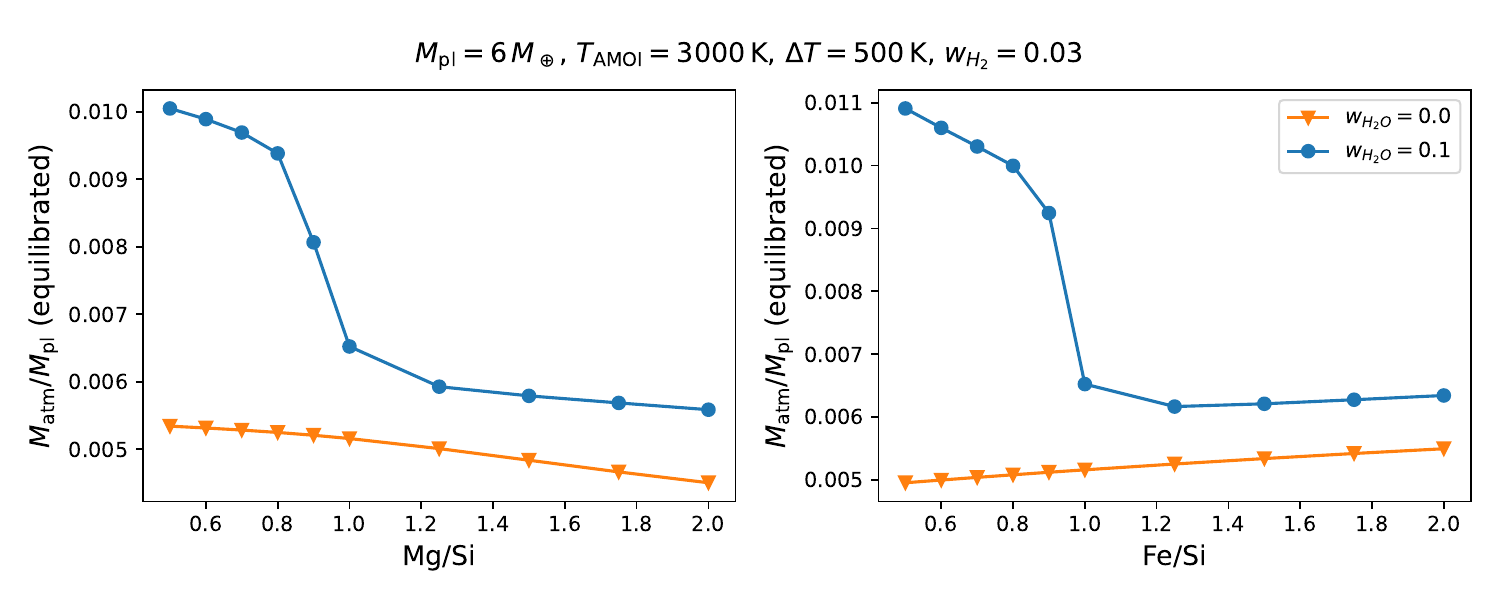}
    \caption{Atmospheric mass fraction as a function of molar Mg/Si (left plot) and Fe/Si (right plot) for a planet with $w_{\mathrm{H}_2\mathrm{O}}=0.0wt\%$ (orange lines) and $w_{\mathrm{H}_2\mathrm{O}}=10wt\%$ (blue lines). }
    \label{fig:Matm_MgSi_FeSi_water}
\end{figure*}

Higher atmospheric mass fractions can be achieved by the accretion of both water and primordial gas (Figure~\ref{fig:Matm_MgSi_FeSi_water}), compared to planets that accreted only primordial gas (Figure \ref{fig:Matm_H2_TAMOI_DeltaT}). In that case, the higher total volatile budget of a planet results in a higher atmospheric mass fraction.
However, for planets with $w_{\ce{H2O}}=10wt\%$, the atmospheric mass fraction also shows a strong dependence on rock composition.
For Mg/Si $\lesssim 1$ or Fe/Si $\lesssim 1$, the atmospheric mass fraction is approximately twice that of an otherwise equivalent dry planet.
For Mg/Si and Fe/Si ratios above unity, the difference between wet and dry planets is more modest.
In contrast, dry planets with $w_{\ce{H2O}}=0.0wt\%$ show only a weak dependence of atmospheric mass fraction on Mg/Si and Fe/Si.

%


\subsection{Atmospheric Metal Mass Fraction}
\label{sec:Zatm}

\begin{figure*}
\centering

\begin{minipage}{0.47\textwidth}
    \centering
    \includegraphics[width=\linewidth]{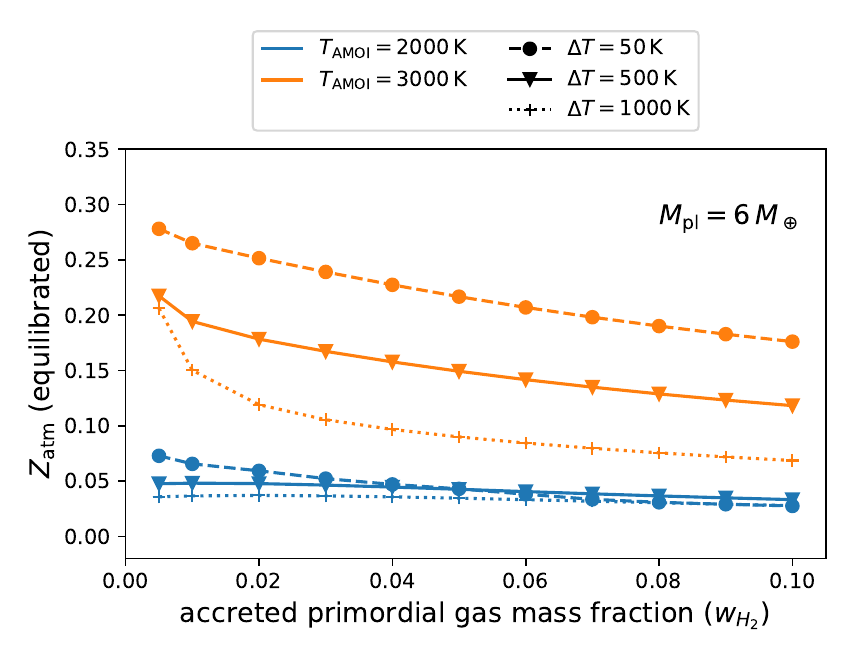}
    \caption{Atmospheric metal mass fraction as a function of the accreted \ce{H2} mass, the temperature at the AMOI (color) and the temperature difference $\Delta T$ between the AMOI and SME (line style). The mass of the planet is $6\,M_\oplus$ and $w_{\ce{H2O}} = 0.0wt\%$.}
    \label{fig:Z_H2_TAMOI_DeltaT}
\end{minipage}
\hfill
\begin{minipage}{0.47\textwidth}
    \centering
    \includegraphics[width=\linewidth]{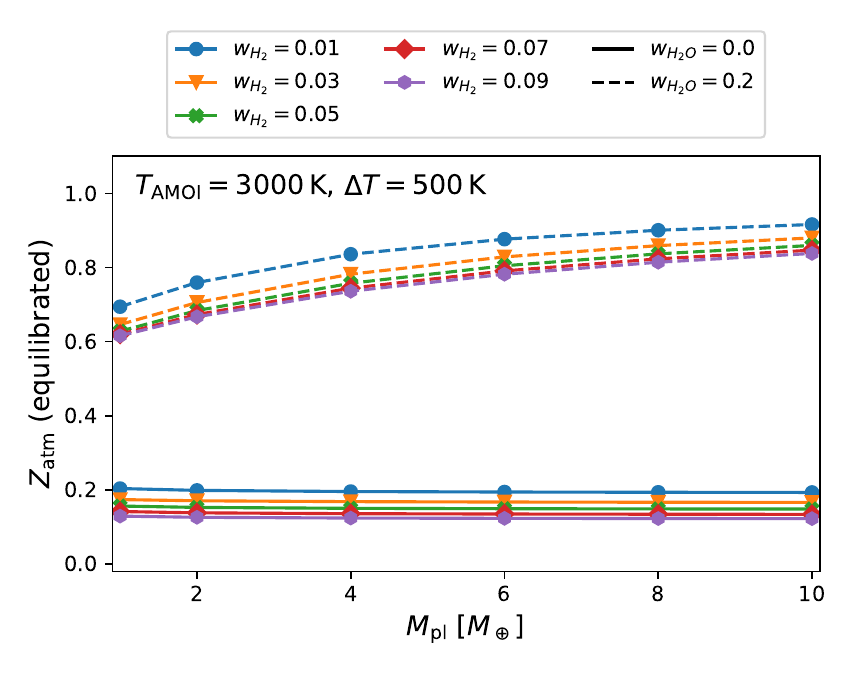}
    \caption{Atmospheric metal mass fraction as a function of the planet mass. Colors \revisedtext{and symbols} show different mass fractions of \ce{H2}. Solid lines correspond to planets with $w_{\mathrm{H_2O}}=0.0wt\%$ and dashed lines to planets with $w_{\mathrm{H_2O}}=20wt\%$. All planets have $T_{\mathrm{AMOI}}=3000\,\mathrm{K}$ and $\Delta T = 500\,\mathrm{K}$.}
    \label{fig:Z_Mpl_wH2_wH2O}
\end{minipage}

\end{figure*}

As illustrated in Figure \ref{fig:Z_H2_TAMOI_DeltaT}, the atmospheric metal mass fraction, which we define as the mass fraction of all species other than \ce{H2} in the gas phase, depends primarily on the thermal state of the planet. 
\revisedtext{We find that hotter interiors yield higher atmospheric metal mass fractions. 
Planets with $T_\mathrm{AMOI}$ of 3000 K have $Z_\mathrm{atm}\sim0.07$–0.3, which corresponds to a factor five to 20 increase compared to the solar metal mass fraction of $Z_\odot = 0.014$ \citep{2021AsplundSun}. 
The atmospheric metal mass fraction of planets with $T_\mathrm{AMOI}=$2000 K is only increased by a factor two to three compared to the solar mass fraction, except for planets with $w_{\ce{H2}}<0.02$ and $\Delta T =$50 K, where the atmospheric metal mass fraction is increased up to a factor of five compared to the solar metal mass fraction.}
Notable, the influence of $\Delta T$ on the atmospheric metal mass fraction is only relevant for hot or young interiors with $T_\mathrm{AMOI}\geq3000\,$K. 
In other words, when metallicities are high, the SME temperature influences the global chemical equilibration state, while its impact is negligible when metallicities are low \revisedtext{(i.e., $\sim 2\times Z_\odot$)}.

The fraction of \ce{H2} in the gas phase scales with the total mass fraction of accreted \ce{H2}.
Consequently, the atmospheric metal mass fraction decreases with increasing $w_{\ce{H2}}$, however the effect is small (Figure \ref{fig:Z_Mpl_wH2_wH2O}). More importantly, the atmospheric metal mass fraction is sensitive to the accreted water mass fraction. 
Planets that have accreted water during formation show higher metallicities by a factor of four to six compared to planets that formed dry (Figure \ref{fig:Z_Mpl_wH2_wH2O}). 
\revisedtext{It is also of note that the atmospheric metal mass }fraction of planets formed dry is independent of the planet mass, while atmospheric metal mass fraction scales positively with planet mass for planets that accreted water.
This trend suggests that the dissolution of oxygen into the planetary interior is sensitive to the pressure at the AMOI, which scales with the planet mass. 

\begin{figure*}
        \centering
    \includegraphics[width=1.0\linewidth]{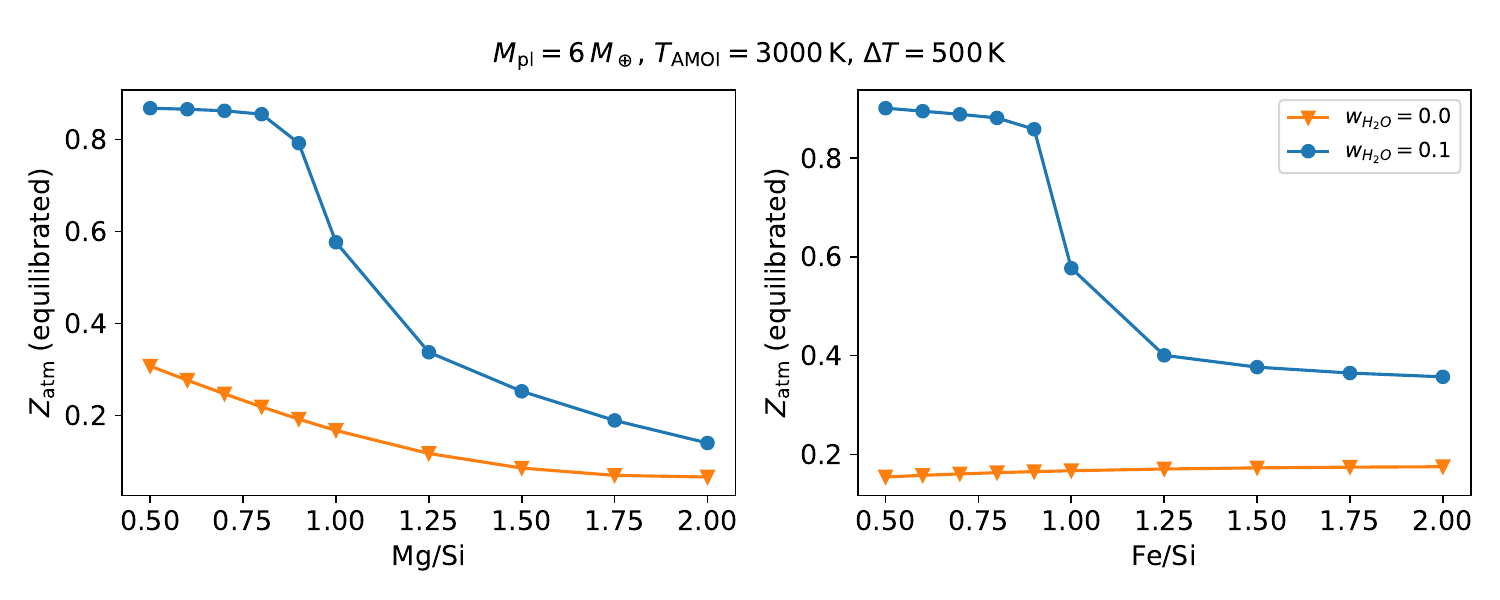}
    \caption{Atmosphere metal mass fraction as a function of molar Mg/Si (left plot) and Fe/Si (right plot) for a planet with $w_{\mathrm{H}_2\mathrm{O}}=0.0wt\%$ (orange lines) and $w_{\mathrm{H}_2\mathrm{O}}=10wt\%$ (blue lines). The total mass is $6\,M_\oplus$,  $TAMOI=3000\,$K and $\Delta T = 500 K$.}
    \label{fig:Z_MgSi_FeSi_water}
\end{figure*}

Finally, Figure \ref{fig:Z_MgSi_FeSi_water} shows the atmospheric metal mass fraction as a function of refractory element ratios for planets formed dry ($w_{\ce{H2O}}=0.0wt\%$) and wet ($w_{\ce{H2O}}=10wt\%$).
Similarly to the atmospheric mass fraction shown in Figure \ref{fig:Matm_MgSi_FeSi_water}, the dependence of atmospheric metal mass fraction on  refractory element composition differs significantly between planets that formed dry or wet.
Planets that formed water-rich show much stronger dependence of their atmospheric metal mass fraction with bulk Mg/Si and Fe/Si ratios. Especially in the ranges of 0.8--1.5 for Mg/Si and 0.9--1.2 for Fe/Si, atmospheric metal mass fraction decreases drastically. The reason is the sudden change in carbon chemistry with methane abundances in the gas phase dropping from log{\ce{CH4}} around -1 to -7 (see Section\ref{sec:abundance}) 
In contrast, for planets formed dry, the atmospheric metal mass fraction decreases smoothly (0.3 - 0.06) with increasing Mg/Si and is independent of Fe/Si. 

\subsection{\revisedtext{Detailed chemical composition of the interior and atmosphere}}
\label{sec:abundance}
\begin{figure*}
    \centering
    \includegraphics[width=0.9\linewidth]{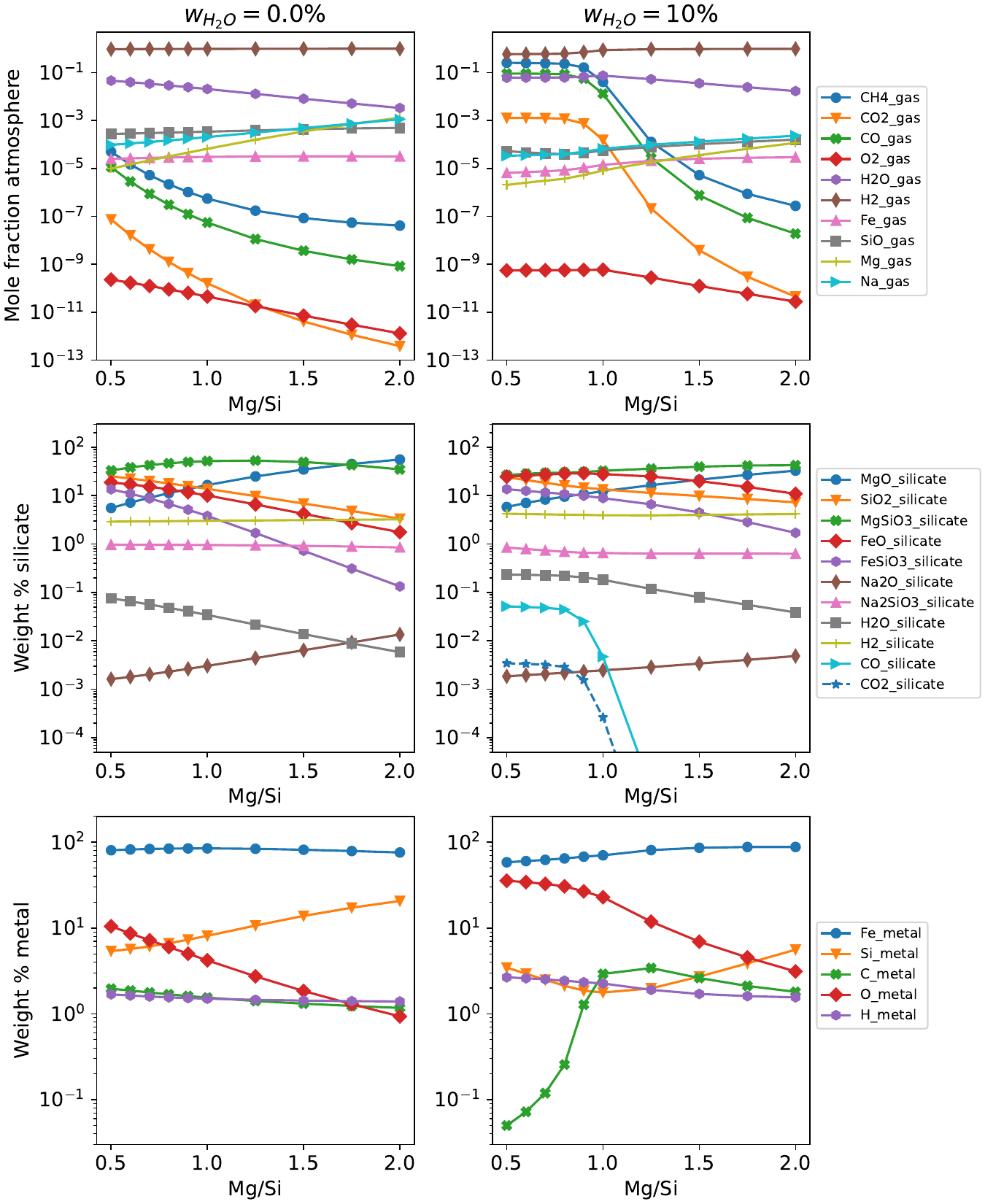}
    \caption{Effect of bulk Mg/Si ratio on chemical composition of the gas (top), silicate (middle), and metallic phase (bottom). The left column shows planets with $w_{\ce{H2O}} = 0.0wt\%$ and the right column planets $w_{\ce{H2O}} = 10wt\%$. For all planets, the total mass is $6\,M_\oplus$, $T_\mathrm{AMOI}=3000\,$K and $\Delta T = 500 K$. The total mass fraction of \ce{H2} is $w=3wt\%$ and the Fe/Si ratio fixed to unity. The composition of the gas phase is given in mole fraction, while the composition of the silicate and metal phases are given in weight \%.}
    \label{fig:chemcompMgSi}
\end{figure*}
\begin{figure*}
    \centering
    \includegraphics[width=0.9\linewidth]{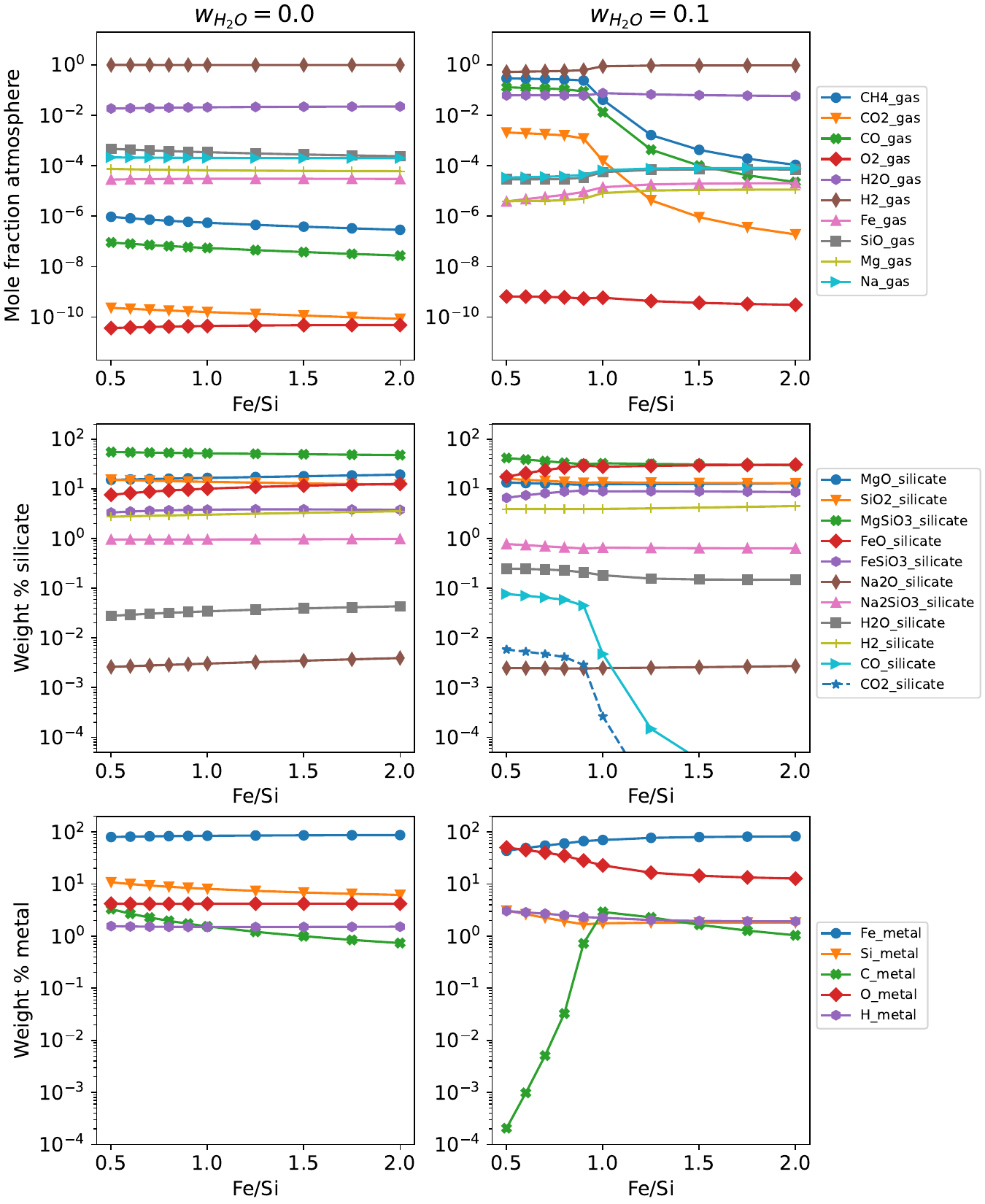}
    \caption{Effect of bulk Fe/Si ratio on chemical composition of the gas (top), silicate (middle), and metallic phase (bottom). The left column shows planets with $w_{\ce{H2O}} = 0.0wt\%$ and the right column planets $w_{\ce{H2O}} = 10wt\%$. For all planets, the total mass is $6\,M_\oplus$, $T_\mathrm{AMOI}=3000\,$K and $\Delta T = 500$\,K. The total mass fraction of \ce{H2} is $w=3wt\%$ and the Mg/Si ratio fixed to unity. The composition of the gas phase is given in mole fraction, while the composition of the silicate and metal phases are given in weight \%. }
    \label{fig:chemcompFeSi}
\end{figure*}

\revisedtext{In order to better understand the influence of the rock composition on the atmospheric properties,} we examine the chemical composition of the gas, silicate, and metallic phases as a function of the refractory element ratios for planets with $w_{\ce{H2O}} = 0.0wt\%$ and $w_{\ce{H2O}} = 10wt\%$ in Figure \ref{fig:chemcompMgSi} and \ref{fig:chemcompFeSi}. 
For all planets, the total planet mass is set to $M_\mathrm{pl} = 6\,M_\oplus$ and the total hydrogen mass fraction to $w_{\ce{H2}}=3wt\%$. The temperature at the AMOI is set to $T_\mathrm{AMOI} = 3000\,$K and $\Delta T = 500\,$K. 

Figure \ref{fig:chemcompMgSi} shows the chemical composition as a function of the molar Mg/Si ratio, with the iron-to-silicate ratio fixed to unity.
The key difference between planets with and without accreted water lies in the dependence on the Mg/Si ratio of the abundance of carbon species in the different phases. 
For dry planets ($w_{\ce{H2O}} = 0.0wt\%$), the majority of carbon is partitioned into the metallic phase independent of the Mg/Si ratio. 
Consequently, C-bearing species are only minor species in the gas and silicate phases. 
In contrast, for planets with $w_{\ce{H2O}} = 10wt\%$ the carbon distribution varies significantly depending on whether the Mg/Si ratio is above or below unity. 
At low Mg/Si, almost all C is in the gas phase and C-bearing species become the most abundance gas species by mole after \ce{H2} resulting in the high metallicities seen in Figure \ref{fig:Z_MgSi_FeSi_water}. 
However, when Mg/Si exceeds unity, the mole fraction of the gaseous C-bearing species drop by five orders of magnitude as C partitions into the metallic phase.
This removal of carbon-bearing gas species causes the drop in atmospheric mass fraction and atmospheric metal mass fraction seen in the left plots of Figure \ref{fig:Matm_MgSi_FeSi_water} and \ref{fig:Z_MgSi_FeSi_water}.
Similarly, the abundance of CO and \ce{CO2} in the silicate phase drops by several orders of magnitude at high Mg/Si ratios. 

Figure \ref{fig:chemcompFeSi} shows the chemical composition as a function of Fe/Si, with the magnesium-to-silicate ratio fixed to unity. 
The compositions of the silicate and gas phases of planets with $w_{\ce{H2O}} = 0.0wt\%$ show only a weak dependence on the Fe/Si, while the fraction of C in the metallic phase decreases with increasing Fe/Si. 
Similar to the dependence on Mg/Si, the distribution of carbon species in planets with $w_{\ce{H2O}} = 10wt\%$ differs significantly depending on whether the Fe/Si ratio is below or above unity. 
For low Fe/Si ($\lesssim 0.9$), \ce{CH4} and CO are the second and third most abundant gas species, raising the atmospheric metal mass fraction as seen in the right plot in Figure \ref{fig:Z_MgSi_FeSi_water}. 
The abundances of C-bearing species in both the gas and silicate phases drop significantly for Fe/Se$\gtrsim1$.
This change in the abundances is caused by a significant increase of C in the metallic phase for higher Fe/Si ratios.

Beyond the dependencies on refractory element budgets, the compositions of both the silicate and metallic phases differ significantly between planets formed with and without water.
For planets with \( w_{\ce{H2O}} = 10wt\% \), the higher total oxygen abundance causes FeO to become the second most abundant species in the silicate phase by weight after \ce{MgSiO3}. In contrast, for dry planets (\( w_{\ce{H2O}} = 0.0wt\% \)), the second most abundant silicate species is either \ce{SiO2} or \ce{MgO}.
The metallic phase also shows a clear distinction. In dry planets, it is dominated by Fe ($\gtrsim 80\,wt\%$), with Si as the second most abundant element. For water-rich planets, however, the mass fraction of O in the metallic phase is strongly enhanced.
These differences in silicate and metallic compositions directly affect atmospheric composition under global chemical equilibrium, since elements partitioned into the metallic and silicate phases are effectively removed from the gas phase.

While water content governs major differences in phase compositions, variations in refractory budgets as Mg/Si and Fe/Si introduce additional systematic changes in silicate and metal phase composition.
We find that increasing the Mg/Si ratio leads to an enrichment of MgO in the silicate phase, independent of the planet’s water content. For dry planets, MgO even becomes the dominant silicate species at Mg/Si $\approx 2$. This enrichment is accompanied by a decrease in oxygen-bearing species in the gas phase, as well as a reduction of elemental oxygen in the metallic phase.
As a result, the declining \ce{H2O} abundance in the gas phase explains the negative trend in $Z_\mathrm{atm}$ with increasing Mg/Si for planets with \( w_{\ce{H2O}} = 0.0wt\% \), as shown in Figure \ref{fig:Z_MgSi_FeSi_water}.
In addition, increasing Fe/Si reduces the oxygen content of the core, making it more iron-dominated. At the same time, the silicate phase becomes increasingly dominated by \ce{MgSiO3}, approaching corresponding mass fractions.
\revisedtext{Overall, varying the bulk refractory composition has a stronger influence on the composition of the atmosphere and interior of water-rich planets. This is because water supplies additional oxygen, which binds differently to Mg, Fe, and Si: Mg and Fe each bind one oxygen atom, whereas Si binds two.}

\subsection{Atmospheric C/O ratios}
\label{sec:CO}
\begin{figure}
    \centering
    \includegraphics[width=1\linewidth]{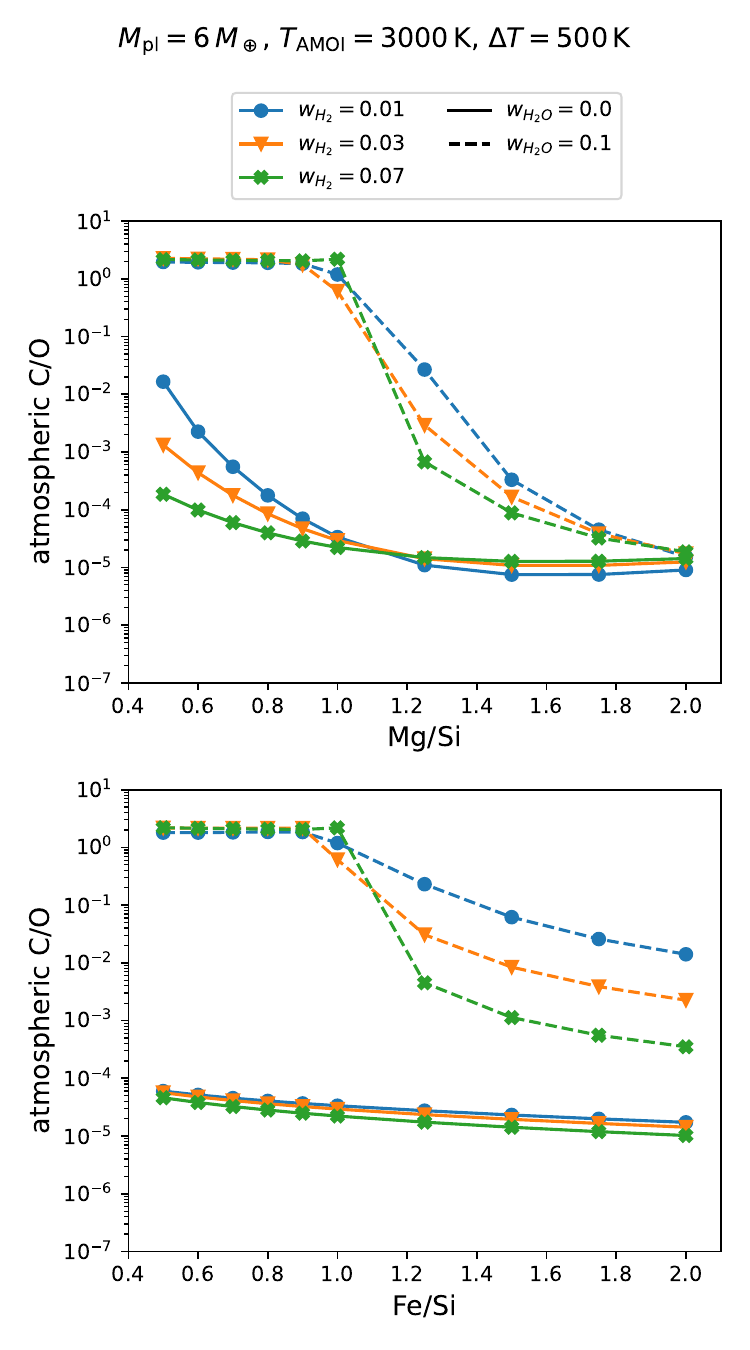}
    \caption{Atmospheric C/O ratio as a function of Mg/Si (top plot) and Fe/Si (bottom plot). The colors show different mass fractions of H$_2$. Solid lines correspond to planets with $w_{\mathrm{H}_2\mathrm{O}}=0.0wt\%$ and dashed lines to planets with $w_{\mathrm{H}_2\mathrm{O}}=10wt\%$.}
    \label{fig:atmosphericCO}
\end{figure}
The global chemical equilibrium framework calculates the composition of the deep atmosphere close to the AMOI. 
The chemical composition of the atmosphere may vary with altitude due effects from vertical mixing, condensation, or photochemistry. 
Nevertheless, \citet{werlen_atmospheric_2025} showed that for strong vertical mixing the deep C/O ratio represents the observable atmospheric C/O.

Figure \ref{fig:atmosphericCO} shows the atmospheric C/O ratio as a function of the rock composition and the mass fraction of accreted \ce{H2} for planets with $w_{\ce{H2O}}=0.0wt\%$ and $w_{\ce{H2O}}=10wt\%$. 
Confirming results from \citet{steinmeyer_coupled_2026}, planets with $w_{\ce{H2O}}=10wt\%$ are characterized by higher atmospheric C/O ratios than their dry counterparts. 

The distinct drop in the abundance of C-bearing species in the gas phase for planets with $w_{\ce{H2O}}=10wt\%$ at Mg/Si and Fe/Si ratios near unity is reflected in the atmospheric C/O ratios.
For Mg/Si $\lesssim 1$ and Fe/Si $\lesssim 1$, the atmospheric C/O ratio remains approximately constant at $\mathrm{C/O}\approx10^0$ and is nearly independent of the total accreted \ce{H2} mass fraction.
For Mg/Si and Fe/Si ratios above unity, the atmospheric C/O ratio decreases as the abundances of Mg and Fe relative to Si increase. 
The magnitude of the decrease associated with increasing Mg/Si ratio exhibits only a  weak dependence on $w_{\ce{H2}}$.
In contrast, the reduction in the atmospheric C/O ratio with increasing Fe/Si ratio clearly scales with $w_{\ce{H2}}$. 
Specifically, the C/O ratio drops to $10^{-2}$ for planets with $w_{\ce{H2}}=1wt\%$ and Fe/Si$=2$, while for planets with $w_{\ce{H2}}=7wt\%$ the C/O ratio is decreased to $10^{-4}$.

For planets formed dry ($w_{\ce{H2O}}=0.0wt\%$ ), the atmospheric C/O ratio shows a much weaker dependence and decreases with increasing Mg/Si ratio until Mg/Si$\approx1.25$, beyond which it is approximately constant. 
In the low Mg/Si regime, atmospheric C/O ratios scale inversely with $w_{\ce{H2}}$, whereas in the high Mg/Si regime the C/O ratios for all three values of $w_{\ce{H2}}$ converge to $\approx 10^{-5}$. The dependence on atmospheric C/O ratio by varying Fe/Si is only marginal for planets formed dry.

\section{Discussion}
\label{sec_discussion}
\subsection{\revisedtext{Influence of hydrogen solubility}}
\label{ssec_influence}
\revisedtext{
The original implementation of the GCE by \citet{schlichting_chemical_2022} assumes that the ratio of dissolved to gaseous \ce{H2} remains constant with pressure (fixed-$k_{\rm D}$ treatment). 
This treatment uses the pressure-independent reference-state Gibbs free energy for dissolved \ce{H2} in the melt from \citet{hirschmann_solubility_2012}.
By construction, it prevents runaway dissolution of \ce{H2} at high pressures.
Recent work on silicate--hydrogen phase equilibria suggests that hydrogen solubility in silicate melts can reach wt\% levels, which is higher than predicted by the fixed-$k_{\rm D}$ treatment \citep{young_differentiation_2025, miozzi_experiments_2025, horn_building_2025, gilmore_coreenvelope_2026}.
We therefore employ a fixed equilibrium constant ($k_{\rm eq}$) treatment in this work, which allows the dissolved \ce{H2} abundance to increase unbounded with pressure.
For this fixed-$k_{\rm eq}$ treatment, the standard-state Gibbs free energy of dissolved \ce{H2} in the melt is computed from a combined fit anchored to the experimental constraints of \citet{hirschmann_solubility_2012} and the melt--gas chemical-potential equivalence from \citet{gilmore_coreenvelope_2026}, following \citet{werlen_effects_2026}. 
A more detailed description of both prescriptions is given in Appendix~\ref{ap:data}.
}

\revisedtext{
The adopted hydrogen-solubility prescription controls how much hydrogen remains in the gas phase versus partitions into the interior. 
For the fixed-$k_{\rm eq}$ prescription, \ce{H2} is highly soluble in silicate melts especially for high temperatures and high $w_{\ce{H2}}$
In contrast, for the fixed-$k_{\rm D}$ prescription, the majority of hydrogen remains in the gas phase. 
As a result, the $k_{\rm eq}$ prescription produces atmospheric mass fractions that are significantly smaller or similar to the fixed-$k_{\rm D}$. 
This leads to systematically lower atmospheric metal mass fractions in the case of the fixed-$k_{\rm D}$ prescription. 
Similarly, the dependence of the atmospheric C/O ratio on $w_{\ce{H2}}$ should be interpreted in light of the hydrogen-solubility prescription adopted in this work.
The overall dependence of the atmospheric C/O ratio on the accreted water mass fraction and Mg/Si ratio found in Section~\ref{sec:CO}, however, is independent of the assumed solubility prescription. 
For a more detailed comparison see Appendix~\ref{app_comp}.
}

\revisedtext{ 
Given the current uncertainties, we stress that the treatment of the hydrogen-solubility remains a key model assumption, which needs to be taken into account when interpreting atmospheric composition trends such as the ones reported in this work. 
Combined with the overall lack of thermodynamic data at higher temperatures and pressures, this highlights the need for further experiments and ab initio simulations to refine our understanding of the atmospheres and interiors of super-Earths and sub-Neptunes.
}

\subsection{\revisedtext{Limitations of the global chemical equilibrium assumption}}
Throughout this work, we assume that the planets are able to reach a state of global chemical equilibrium. 
However, the development of layered convection within the interior and atmosphere may prevent the planet from reaching this state by suppressing the mixing of the different phases. 
Specifically, mean molecular weight gradients can inhibit convection in the atmospheres of sub-Neptunes\citep[e.g.,][]{2017Leconte,misener_importance_2022,2022Markham,2024Leconte3d}. 
The emergence of a radiative region in the deep atmosphere creates a barrier between the atmosphere-magma ocean interface and the upper atmosphere. 
This barrier potentially prevents outgassed species from reaching the upper, observable atmosphere.
Furthermore, the internal dynamics of magma oceans in super-Earths and sub-Neptunes remain poorly understood. 
Previous research suggest that layered or sluggish convection may occur in super-Earths \citep[e.g.,][]{2012Stamenkovicinfluence,2020Spaargaren}, while other studies find evidence for vigorous convective mixing within silicate melts \citep[e.g.,][]{lichtenberg_redox_2021}.
Consequently, future work is required to determine under which circumstances large scale mixing is enabled or suppressed in super-Earths and sub-Neptunes.
 
The process of core-mantle differentiation significantly influences the chemical evolution of a planet by decoupling the metallic phase from the silicate and gas phases. 
The partitioning of volatiles into the metallic phase during core formation permanently excludes these species from subsequent global chemical equilibrium reactions \citep{schlichting_chemical_2022, luo_interior_2024}. 
Consequently, in differentiated planets, the chemical equilibrium is limited to reactions between the silicate and gas phases. 
However, recent works highlighted that volatile-rich planets might not possess a differentiated metal core \citep{2025Huanglimits,young_differentiation_2025,young_influences_2026,luo2026}. More work is therefore needed to understand the interior structure of such planets.

\revisedtext{In addition, we do not explicitly account for possible H$_2$--H$_2$O phase separation within the atmosphere. Our framework treats the gas as a single homogeneous phase. However, for the AMOI temperatures considered here, $T_{\rm AMOI}=2000$--$3000\,{\rm K}$, the phase diagram of \citet{gupta_miscibility_2025} indicates that H$_2$--H$_2$O mixtures remain miscible up to pressures of order $10$--$30\,{\rm GPa}$, depending on composition. In the fixed-$k_{\rm eq}$ models adopted for our main results, the AMOI pressures remain below this range because efficient dissolution of hydrogen limits the atmospheric mass. We therefore do not expect H$_2$--H$_2$O demixing to affect the gas phase at the AMOI for the parameter space explored here. Nevertheless, such demixing may become relevant for cooler planets, higher atmospheric masses, or hydrogen-solubility prescriptions that leave more hydrogen in the atmosphere.}

The thermodynamic data needed to calculate the global equilibrium is taken from a diverse suite of sources, combining both laboratory data and results from ab initio simulations, see Appendix \ref{ap:data}. 
However, the lack of data at high temperatures and pressures as well as volatile-rich conditions remains the major limitation of this framework as it limits the parameter space that can be explored. \revisedtext{This limitation is especially important for young or highly irradiated sub-Neptunes, for which the atmosphere--magma ocean interface temperature may exceed $3000\,{\rm K}$. In addition, pressure-dependent thermodynamic data are generally not available over the relevant parameter range, and pressure effects are therefore neglected for many species in the present implementation (Appendix~\ref{ap:gce}). Calculations outside the nominal temperature range of the adopted thermodynamic dataset should therefore be interpreted as extrapolations rather than fully validated predictions.} We therefore recommend the use of our framework for \revisedtext{AMOI} temperature ranges between 2000 and 3000 K, and hydrogen mass fractions $w_{\ce{H2}}<10wt\%$.
\revisedtext{In addition to the lack of thermodynamic data at high temperatures, recent works suggest that for AMOI temperatures exceeding $3000\,{\rm K}$ hydrogen and silicates become miscible as disscued in the next paragraph \citep{gilmore_coreenvelope_2026}. Regarding a lower AMOI temperature limit, the GCE model becomes invalid when a magma ocean solidifies and ceases to exists which, depending on rock composition, can be initiated when AMOI temperatures fall well below $2000\,{\rm K}$.}

While the global chemical equilibrium framework presented in this work focuses on the chemical interactions between the metal, silicate, and gas phase, future models should further account for miscibility effects. 
Several studies indicate that the interiors of sub-Neptune can reach pressure-temperature conditions where the phases become miscible, resulting in a single, homogeneous mixture \citep{2022Markham,young_phase_2024,young_differentiation_2025,rogers_redefining_2025,gilmore_coreenvelope_2026}.
Incorporating these phase transitions is critical, as miscibility can increase the amount of volatiles dissolved in the interior.

\subsection{\revisedtext{Compositional correlation between stars and planets}}
A commonly adopted assumption is that the relative abundances of these refractory, rock-forming elements (Mg, Si, Fe) are largely preserved during planet formation in protoplanetary disks. This is supported by the broadly similar elemental ratios observed between primitive chondrites and the solar photosphere \citep{Palme_solar_2014}, which underpin estimates of the bulk silicate Earth \citep{mcdonough_composition_1995}. In addition, condensation models predict that key ratios such as Mg/Si and Fe/Si remain similar between stars, disks, and planets \citep{bond_compositional_2010, carter-bond_compositional_2012, moriarty_chemistry_2014, thiabaud_elemental_2015,jorge_forming_2022}. Consequently, stellar refractory abundances are often used as proxies for planetary compositions \citep{dorn_can_2015, dorn_generalized_2017, unterborn_inward_2018}.

However, observational tests of the star–planet compositional connection yield mixed results. Some studies find strong correlations \citep{adibekyan_compositional_2021}, while others find weak or no clear link \citep{brinkman_revisiting_2024}, with recent studies reporting a wide range of outcomes, from weak to moderate correlations, often depending on sample selection and modelling assumptions \citep{santos_constraining_2015,santos_constraining_2017,plotnykov_chemical_2020,schulze_probability_2021,brinkman_compositions_2025, behmard_link_2025}. Evidence from polluted white dwarfs provides tentative support for a connection \citep{bonsor_host-star_2021, aguilera-gomez_host_2025,rogers_silicate_2025}, though uncertainties remain \citep{brouwers_asynchronous_2023}. Overall, planetary refractory element ratios are likely inherited to first order, but the strength of this link remains uncertain. 

\section{Conclusion}
\label{sec_conclusion}
This paper investigates the influence of bulk composition and thermal state on the atmospheres of Earth- to sub-Neptune-sized planets, utilizing the global chemical equilibrium framework first presented in \citet{schlichting_chemical_2022}.
We conduct a large parameter study varying the mass fractions of accreted \ce{H2}-dominated primordial gas ($w_{\ce{H2}}$) and water ($w_{\ce{H2O}}$), as well as the bulk refractory composition using an updated implementation of the global chemical equilibrium first presented in \citet{schlichting_chemical_2022}. Our updated implementation has an increased computational efficiency by more than two orders of magnitude thanks to various improvements of using a gradient-based optimizer and solving for molar fraction in log-space (details see Appendix \ref{ap:gce}).
\revisedtext{In addition, the treatment of hydrogen solubility has been updated to reflect recent works on silicate-hydrogen phase equilibria, which suggest that the solubility of hydrogen in silicate melts is higher than predicted by the original treatment used in \citet{schlichting_chemical_2022}.}

Our results demonstrate that \revisedtext{for the updated hydrogen solubility treatment} the temperature at the atmosphere-magma ocean interface ($T_\mathrm{AMOI}$) and the initial water budget are the main factors determining the atmospheric mass fraction and chemical composition. 
The atmospheric mass fraction decreases with increasing $T_\mathrm{AMOI}$ due to the enhanced dissolution of volatiles into the planetary interior. 
In contrast, the atmospheric metal mass fraction exhibits a strong positive correlation with the thermal state of the planet. 
The addition of water raises both the atmospheric mass fraction and atmospheric metal mass fraction compared to a planet formed dry. 
In contrast, the initial \ce{H2} mass fraction plays only a secondary role because most hydrogen dissolves into the interior of the planet. 
 
For planets with $w_{\ce{H2O}}>0.0wt\%$, the refractory molar ratios (Mg/Si and Fe/Si) in the considered rock material significantly influence the distribution of carbon across the different phases. 
For Mg/Si and Fe/Si ratios below unity, the majority of carbon is stored in gaseous species, leading to a marked increase in atmospheric mass fraction and atmospheric metal mass fraction compared to planets with refractory element ratios above unity. 
For planets formed dry ($w_{\ce{H2O}}=0.0wt\%$), on the contrary, these refractory element ratios play only a minor role for the atmosphere composition. 

Thanks to the JWST, we are now able to characterize the atmospheres of sub-Neptunes in detail. 
Already, the first new observations revealed a large diversity in their atmospheric composition, from \ce{H2}/He-dominated atmospheres with solar atmospheric metal mass fraction \citep{Davenport_toi421b_2025} to high-atmospheric metal mass fraction or even water-dominated atmospheres \citep[e.g.,][]{kempton_reflective_2023,beatty_sulfur_2024,benneke_jwst_2024,piaulet_GJ9827d_2024}.
\revisedtext{Ongoing efforts aim to connect these observations to interior--atmosphere models \citep[see e.g.,][]{nixon_magma_2025,werlen_effects_2026,dorn_atmospheric_2026}.}
With the launch of the Ariel mission \citep{tinetti_ariel_2021} and ESO's Extremely Large Telescope \citep{2023ELT}, the number of well characterized sub-Neptune atmospheres spanning a wide range of equilibrium temperatures will increase significantly. 

Based on our findings, we suggest that future observational surveys should strategically target sub-Neptunes orbiting host stars with a wide range of Mg/Si and Fe/Si ratios to constrain their formation histories. 
A significant correlation between the host refractory element ratios and the atmospheric C/O ratio would serve as a robust indicator that sub-Neptunes form beyond the water ice line. 
Conversely, low atmospheric C/O ratios and a weak dependence on the stellar composition would signify formation inside the water ice line. Current estimates of C/O based on JWST-observations of sub-Neptunes \citep{benneke_jwst_2024,beatty_sulfur_2024,Davenport_toi421b_2025,schmidt_comprehensive_2025,fernandez-rodriguez_atmospheric_2025,felix_competing_2025} suggest some diversity of C/O including solar or super-solar values that is best explained by formation beyond the water ice-line. If this is true, sub-Neptunes orbiting host stars with high Mg/Si or Fe/Si ratios should show carbon-depleted and \ce{H2}-\ce{H2O} dominant atmospheres; assuming that the star's Mg/Si or Fe/Si is a good first order proxy for the planetary ratios.

Furthermore, observations of sub-Neptunes should ideally span a wide range of planetary ages. 
Younger sub-Neptunes with higher $T_\mathrm{AMOI}$, are expected to result in higher atmosphere metallicities and smaller atmospheric mass fractions compared to older, more evolved sub-Neptunes as a result of the dissolution of volatiles into the interior \citep[see also][]{steinmeyer_coupled_2026}. 
However, any interpretation of atmospheric mass fractions across different evolutionary stages must account for the effect of atmospheric mass loss. 
\revisedtext{In addition, the preferable loss of lighter elements leads to an increase in the atmospheric metal mass fraction with time, especially for planets close to the radius valley \citep{2025Cherubim,2025LoucaMiguel,2026Valatsou}. 
To interpret population level trends with planetary ages, we therefore need models that account for both the chemical interaction between the atmosphere and interior and fractionated mass loss.} 

This work demonstrates that the geochemical interactions between the atmosphere and the underlying interior need to be taking into account when analyzing the observed spectra of Earth- to sub-Neptune sized planets. 
While the atmospheric C/O ratio of sub-Neptunes has been proposed as a potential probe of their bulk volatile composition \citep{werlen_atmospheric_2025,steinmeyer_coupled_2026}, our results indicate that this ratio is further influenced by the refractory composition as well. 
Consequently, stellar abundance constraints are indispensable for interpreting the atmospheric signatures of these planets. 

\begin{acknowledgments}

C.D acknowledges support from the Swiss National Science Foundation under grant TMSGI2\_211313. H.E.S gratefully acknowledges support from NASA under grant No. 80NSSC18K0828. E.D.Y. acknowledges support from NASA grant No. 80NSSC21K0477 issued through the Emerging Worlds program. This work has been carried out within the framework of the NCCR PlanetS supported by the Swiss National Science Foundation under grant 51NF40\_205606. We acknowledge the use of large language models (LLMs) to improve the grammar, clarity, and readability of the manuscript.

 \end{acknowledgments}

\begin{contribution}
S.L.G., M.-L.S., and A.W. contributed equally to this work and share first authorship. C.D. conceived and coordinated the entire study. S.L.G. and A.W. contributed to the development, testing, and publication of the Global Chemical Equilibrium (GCE) code and co-wrote the methods section. M-L.S. and C.D. developed the scientific motivation and overarching science case for this study, and M.-L.S. produced the figures for the main text. H.E.S. and E.D.Y. developed the original global equilibrium framework and its thermodynamic foundations and contributed to the writing. All authors read and provided comments for the manuscript.
\end{contribution} 

\section*{ORCID iDs}

\noindent 
Simon L. Grimm \orcidlink{0000-0002-0632-4407} \href{https://orcid.org/0000-0002-0632-4407}{0000-0002-0632-4407} \\
Marie-Luise Steinmeyer \orcidlink{0000-0003-0605-0263} \href{https://orcid.org/0000-0003-0605-0263}{0000-0003-0605-0263} \\
Aaron Werlen \orcidlink{0009-0005-1133-7586} \href{https://orcid.org/0009-0005-1133-7586}{0009-0005-1133-7586} \\
Caroline Dorn \orcidlink{0000-0001-6110-4610} \href{https://orcid.org/0000-0001-6110-4610}{0000-0001-6110-4610} \\
Hilke E. Schlichting \orcidlink{0000-0002-0298-8089} \href{https://orcid.org/0000-0002-0298-8089}{0000-0002-0298-8089} \\
Edward D. Young \orcidlink{0000-0002-1299-0801} \href{https://orcid.org/0000-0002-1299-0801}{0000-0002-1299-0801}\\
 
\FloatBarrier
\appendix
\twocolumngrid

\section{Global Chemical Equilibrium Framework}
\label{ap:gce}

\subsection{Reaction Network}
\label{ap:chem_network}
We present here the example reaction network used throughout this study. This network was previously applied in \cite{werlen_atmospheric_2025,werlen_sub-neptunes_2025} and extends the original framework presented in \cite{schlichting_chemical_2022} by including carbon partitioning into the metal phase.

The default configuration consists of three phases (metal, silicate, gas), seven elements (Si, Mg, O, Fe, H, Na, C), and 26 species and 19 linearly independent reactions.

\noindent\textbf{Silicate species}\\
\ce{MgO}, \ce{SiO2}, \ce{MgSiO3}, \ce{FeO}, \ce{FeSiO3}, 
\ce{Na2O}, \ce{Na2SiO3}, 
\ce{H2}, \ce{H2O}, \ce{CO}, \ce{CO2}.

\noindent\textbf{Metal species}\\
\ce{Fe}, \ce{Si}, \ce{O}, \ce{H}, \ce{C}.

\noindent\textbf{Gas species}\\
\ce{H2}, \ce{CO}, \ce{CO2}, \ce{CH4}, \ce{O2}, \ce{H2O}, 
\ce{Fe}, \ce{Mg}, \ce{SiO}, \ce{Na}.

\noindent\textbf{Independent reactions}

\begin{align}
\mathrm{Na_2SiO_3}\,(\mathrm{sil}) \rightleftharpoons \mathrm{Na_2O}\,(\mathrm{sil}) + \mathrm{SiO_2}\,(\mathrm{sil}) \tag{R1} \\
\tfrac{1}{2}\,\mathrm{SiO_2}\,(\mathrm{sil}) + \mathrm{Fe}\,(\mathrm{met}) \rightleftharpoons \mathrm{FeO}\,(\mathrm{sil}) + \tfrac{1}{2}\,\mathrm{Si}\,(\mathrm{met})  \tag{R2} \\
\mathrm{MgSiO_3}\,(\mathrm{sil}) \rightleftharpoons \mathrm{MgO}\,(\mathrm{sil}) + \mathrm{SiO_2}\,(\mathrm{sil})  \tag{R3} \\
\mathrm{O}\,(\mathrm{met}) + \tfrac{1}{2}\,\mathrm{Si}\,(\mathrm{met}) \rightleftharpoons \tfrac{1}{2}\,\mathrm{SiO_2}\,(\mathrm{sil})  \tag{R4} \\
2\,\mathrm{H}\,(\mathrm{met}) \rightleftharpoons \mathrm{H_2}\,(\mathrm{sil})  \tag{R5} \\
\mathrm{FeSiO_3}\,(\mathrm{sil}) \rightleftharpoons \mathrm{FeO}\,(\mathrm{sil}) + \mathrm{SiO_2}\,(\mathrm{sil})  \tag{R6} \\
2\,\mathrm{H_2O}\,(\mathrm{sil}) + \mathrm{Si}\,(\mathrm{met}) \rightleftharpoons \mathrm{SiO_2}\,(\mathrm{sil}) + 2\,\mathrm{H_2}\,(\mathrm{sil})  \tag{R7} \\
\mathrm{CO}\,(\mathrm{gas}) + \tfrac{1}{2}\,\mathrm{O_2}\,(\mathrm{gas}) \rightleftharpoons \mathrm{CO_2}\,(\mathrm{gas})  \tag{R8} \\
\mathrm{CH_4}\,(\mathrm{gas}) + \tfrac{1}{2}\,\mathrm{O_2}\,(\mathrm{gas}) \rightleftharpoons 2\,\mathrm{H_2}\,(\mathrm{gas}) + \mathrm{CO}\,(\mathrm{gas})  \tag{R9} \\
\mathrm{H_2}\,(\mathrm{gas}) + \tfrac{1}{2}\,\mathrm{O_2}\,(\mathrm{gas}) \rightleftharpoons \mathrm{H_2O}\,(\mathrm{gas})  \tag{R10} \\
\mathrm{FeO}\,(\mathrm{sil}) \rightleftharpoons \mathrm{Fe}\,(\mathrm{gas}) + \tfrac{1}{2}\,\mathrm{O_2}\,(\mathrm{gas})  \tag{R11} \\
\mathrm{MgO}\,(\mathrm{sil}) \rightleftharpoons \mathrm{Mg}\,(\mathrm{gas}) + \tfrac{1}{2}\,\mathrm{O_2}\,(\mathrm{gas})  \tag{R12} \\
\mathrm{SiO_2}\,(\mathrm{sil}) \rightleftharpoons \mathrm{SiO}\,(\mathrm{gas}) + \tfrac{1}{2}\,\mathrm{O_2}\,(\mathrm{gas})  \tag{R13} \\
\mathrm{Na_2O}\,(\mathrm{sil}) \rightleftharpoons 2\,\mathrm{Na}\,(\mathrm{gas}) + \tfrac{1}{2}\,\mathrm{O_2}\,(\mathrm{gas})  \tag{R14} \\
\mathrm{H_2}\,(\mathrm{gas}) \rightleftharpoons \mathrm{H_2}\,(\mathrm{sil})  \tag{R15} \\
\mathrm{H_2O}\,(\mathrm{gas}) \rightleftharpoons \mathrm{H_2O}\,(\mathrm{sil})  \tag{R16} \\
\mathrm{CO}\,(\mathrm{gas}) \rightleftharpoons \mathrm{CO}\,(\mathrm{sil})  \tag{R17} \\
\mathrm{CO_2}\,(\mathrm{gas}) \rightleftharpoons \mathrm{CO_2}\,(\mathrm{sil})  \tag{R18} \\
\mathrm{C}\,(\mathrm{met}) + \mathrm{O}\,(\mathrm{met}) \rightleftharpoons \mathrm{CO}\,(\mathrm{sil})  \tag{R19}
\end{align}

The reaction set spans the full reaction space of the adopted system. Any linear combination of the reactions listed above is inherently included in the equilibrium formulation. The network therefore represents one possible basis of the reaction space and is not unique.

With 19 independent reactions, 7 elemental mass balance constraints, and 3 phase normalization constraints, the system consists of 29 coupled nonlinear equations (see Section~\ref{sec_chemicalmodel}). These are solved simultaneously for the mole fractions $x_{i,k}$ of the 26 species and the total mole numbers $N_k$ of the three phases.

Because the equilibrium conditions involve $\ln(x_{i,k})$, and mole fractions satisfy $0 < x_{i,k} < 1$ and may become extremely small, directly solving for $x_{i,k}$ can lead to numerical instability. We therefore adopt $\ln(x_{i,k})$ as the independent variables during optimization and transform the solution back to linear space after convergence.

\subsection{Solving the Governing Equations}
\label{AppendixSolver}

We solve the 30 equations described before by minimizing a cost function defined as
\begin{equation}
    \label{eq_ctot}
    \text{CTOT} = \sum_{i = 1}^{20} f_i ^2 + \sum_{i = 1}^7 m_i^2 + \sum_{i = 1}^3 s_i^2,
\end{equation}
containing the 19 reaction equations $f_i = 0$, the seven mass balance equations $m_i = 0$, and the three fraction equations $s_i = 0$. We use the ADAM optimizer \citep{kingma+2017} to minimize the cost function. This is a self-learning gradient descent method that requires the derivatives of Equation \ref{eq_ctot} with respect to the 30 variables used. These derivatives can include many terms and be numerically expensive to calculate. Therefore, it is essential to optimize them numerically as much as possible to reduce the run time of the code. It is also of great importance that this optimization step is automated and that no manual optimization is required because the code must be flexible enough to handle modifications on the chemical network used. 

To fulfill these needs, we split the entire build workflow into several steps. In the first step, the user can modify the network equations as needed. In order to keep this simple, the equations are implemented in a Python file, which can be modified easily. We then use the Python symbolic mathematics package SymPy\footnote{https://www.sympy.org} \citep{sympy} to automatically calculate all the required derivatives.

Since these obtained derivatives terms can be very long and computationally expansive to calculate, we Simplify them by using the SymPy Common Subexpression Elimination (CSE) technique. \revisedtext{This procedure does not approximate or alter the symbolic derivatives. Instead, repeated subexpressions are stored as temporary variables and evaluated only once, improving computational efficiency without changing the mathematical form of the derivatives.} This step significantly reduces the amount of computational operations involved, and therefore speeds up the overall computation time. 

In the last step of the building phase, we automatically translate the obtained Python code to C++ code to further improve performance. 

All the described steps are implemented into a Makefile and are executed automatically.

\subsection{Solver Strategies}

The ADAM optimizer used is an adaptive gradient descent method that minimizes a particular cost function. Tests have shown that this solver works very well in most cases, and the global minimum can be found in a fraction of a second. However, it can happen that the solver gets stuck in a local minimum and misses the global solution. In order to prevent this, multiple walkers can be used to cover a larger parameter space. Even using 100 walkers with random initial conditions is fast enough for most applications, and the chances that all walkers miss the global solution are very low. However, to further improve the quality of the solution, the solver supports an iterative process, where after each iteration, the walkers are redistributed around the best solution from the previous run. 

This described strategy is a good balance between simplicity and performance. In principle, there would be more advanced solvers available, like the Stein Variational Gradient Descent (SVGD) method \citep{LiuWang2019} where the different walkers are combined to scan a large parameter space. Tests have shown that, in most cases, such a more advanced method is not needed for our problem set, but future versions of the code could include them.

\subsection{Performance and Example Runs}
\label{sect_example}

\begin{figure}
    \centering
    \includegraphics[width=1\linewidth]{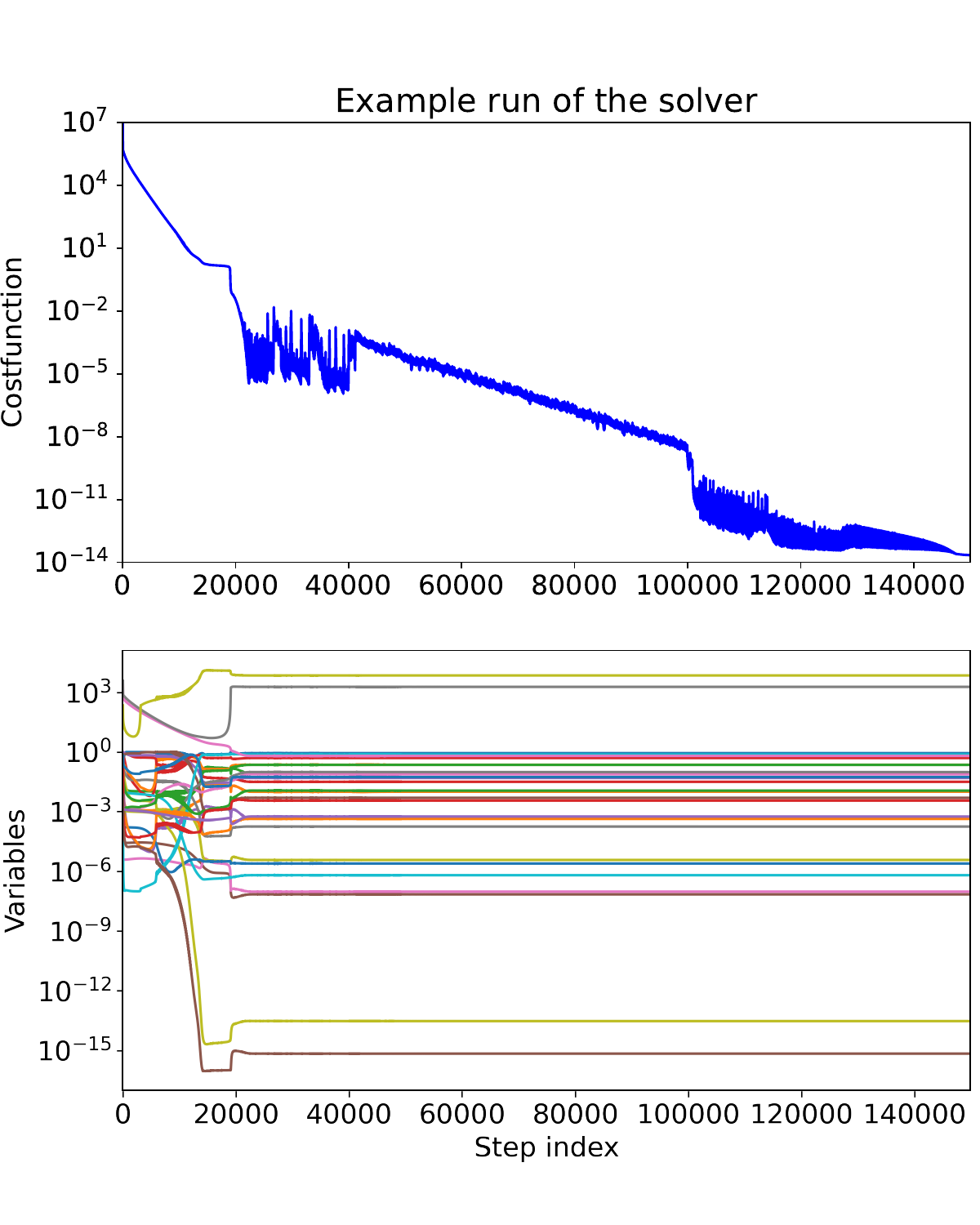}
    \caption{Top panel: Evolution of the cost function with the iteration index. \\
    Bottom panel: Evolution of the model parameters with the iteration index. The mole fractions of species $i$, $x_i$, are bound between 0 and 1, the three numbers of moles of phase $k$, $N_k$, can have values greater than 1.\\
    The calculation of the shown 150000 steps takes 0.07 seconds on a 5.7 GHz AMD Ryzen 9 7950X3D CPU.}
    \label{fig:run1}
\end{figure}

\begin{figure}
    \centering
    \includegraphics[width=1\linewidth]{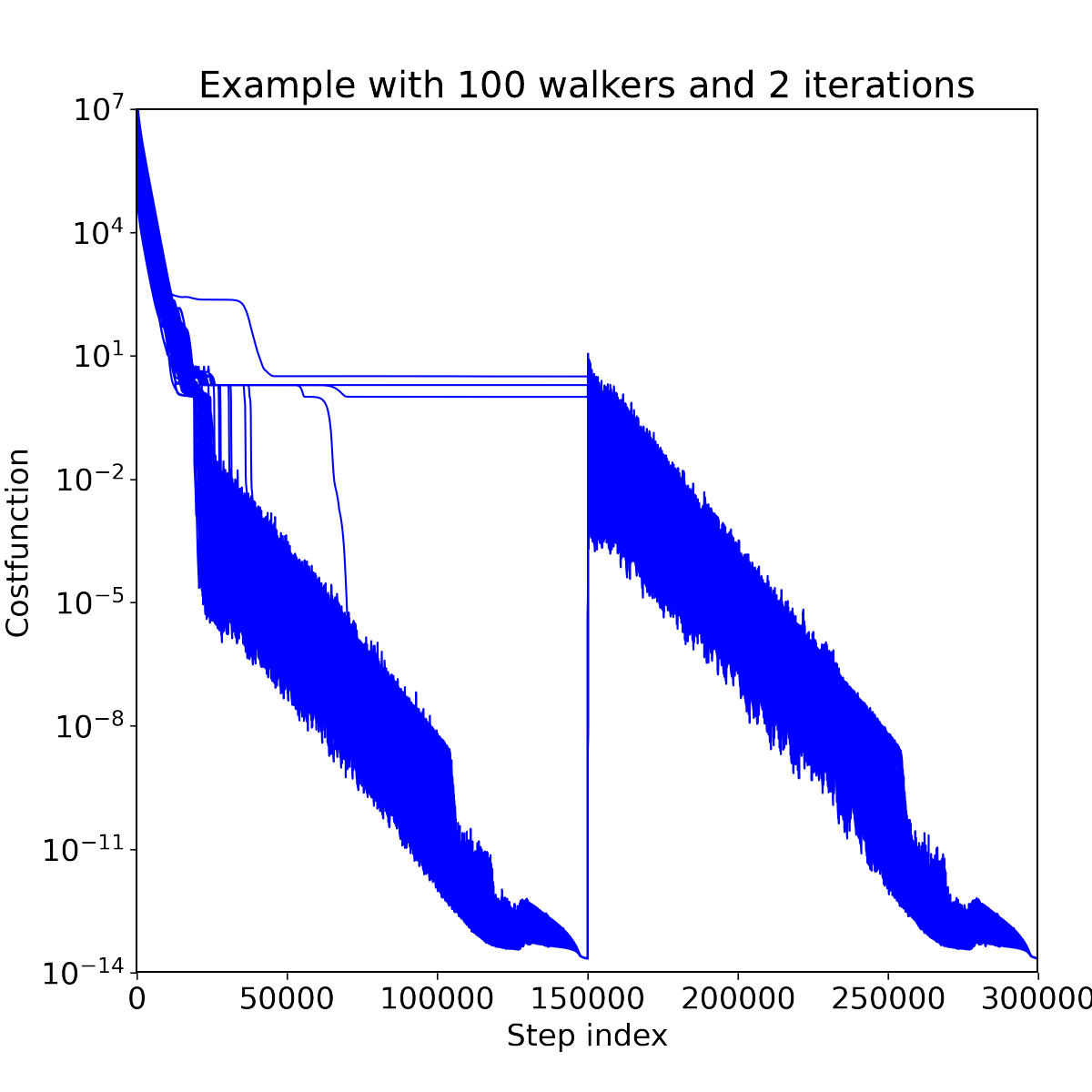}
    \caption{Same initial conditions as shown in Figure \ref{fig:run1} but with 100 walkers and two iterations of each 150000 steps. During the first iteration, 3 walkers out of 100 are stuck in a local minimum. After redistributing the initial conditions in the second iteration, all 100 walker find the same solution.\\
    The calculation of the shown 2 $\times$ 150000 steps with 1000 walkers takes 12.2 seconds on a 5.7 GHz AMD Ryzen 9 7950X3D CPU.}
    \label{fig:run100}
\end{figure}

\begin{table}[t]
\centering
\caption{Initial conditions of the example run shown in Figures~\ref{fig:run1} and \ref{fig:run100}. Listed are the number of moles of the elements used, followed by the planet mass, the AMOI temperature, and the SME temperature.}
\label{tab:IC}
\begin{tabular}{lcl}
\hline
$n_{\mathrm{Si}}$ &=& 1802.90 \\
$n_{\mathrm{Mg}}$ &=& 1838.73 \\
$n_{\mathrm{O}}$  &=& 5854.04 \\
$n_{\mathrm{Fe}}$ &=& 1704.05 \\
$n_{\mathrm{H}}$  &=& 355.48 \\
$n_{\mathrm{Na}}$ &=& $10^{-3}$ \\
$n_{\mathrm{C}}$  &=& 0.11 \\
$M_{\rm \color{blue}{pl}}$             &=& 0.50499144 $M_\oplus$ \\
$T_{\mathrm{AMOI}}$ &=& 4000 K \\
$T_{\mathrm{SME}}$  &=& 4500 K \\
\hline
\end{tabular}
\end{table}

In order to demonstrate how the implemented solver performs, we show here the full evolution of an example run.  The chemical initial conditions used are summarized in Table \ref{tab:IC}. Figure \ref{fig:run1} shows in the top panel the evolution of the cost function CTOT (Equation \ref{eq_ctot}) and in the bottom panel the evolution of all involved model parameters, which include the mole fractions of all species and the number of moles in each phase. While the solver reaches a CTOT value of less than 1 in $\approx$ 20000 steps, it takes 150000 steps to reach a value of $10^{-15}$.  In Figure \ref{fig:run100} the same example is shown with using 100 walkers and 2 iteration steps of each 150'000 steps. At the end of the first iteration, three walkers are stuck at a local minimum and are unable to find the best solution. For the second iteration, the 100 walkers are redistributed around the best value, to possibly still refine the global solution.

\revisedtext{It is important to note that not all initial conditions lead to such low $C_{\rm TOT}$ values of $10^{-15}$. Solutions with large residuals, for example $C_{\rm TOT}\sim0.1$--$1.0$, should not be trusted without further checks. Since $C_{\rm TOT}$ is a residual metric rather than a formal mathematical convergence guarantee, convergence should be assessed by verifying that the residual is sufficiently small and that repeated runs with different solver hyperparameters, such as the number of walkers, the number of iterations, or the initial conditions, recover the same solution.}

\subsection{Thermodynamic Data}
\label{ap:data}

The thermodynamic data adopted in this work follow the compilation given in the Appendix of \citet{schlichting_chemical_2022}, with the addition of carbon partitioning between the silicate and metal phases based on \citet{blanchard_metalsilicate_2022} and implemented as described in \citet{werlen_atmospheric_2025}.

Hydrogen solubility in the silicate melt is implemented using two alternative prescriptions. 
The ambiguity arises from the incomplete characterization of how the Gibbs free energy for hydrogen partitioning varies with pressure and temperature. 
The relevant partitioning reaction is
\begin{equation}
    \ce{H2_{,\mathrm{g}} \rightleftharpoons H2_{,\mathrm{s}}}
    \label{eqn:H2solubility}
\end{equation}

In the first treatment, following the original implementation of \citet{schlichting_chemical_2022}, the dissolved-to-gaseous hydrogen ratio is fixed through
\begin{equation}
    k_{\rm D}=
    \frac{x_{\mathrm{H_2},\mathrm{s}}}{x_{\mathrm{H_2},\mathrm{g}}}
    =
    \exp\left(-\frac{\Delta \hat{G}^{\rm ref}}{RT}\right),
    \label{eqn:kdH2}
\end{equation}
where the superscript ``ref'' denotes the free energy at the experimental reference pressure. 
This formulation uses the pressure-independent reference-state Gibbs free energy from \citet{hirschmann_solubility_2012}. 
By construction, $x_{\mathrm{H_2},\mathrm{s}}/x_{\mathrm{H_2},\mathrm{g}}$ remains constant with pressure, preventing runaway dissolution of \ce{H2} at high pressures.

In the second treatment, the equilibrium constant is instead kept fixed:
\begin{equation}
    k_{\mathrm{eq}} =
    \frac{a_{\mathrm{H_2},\mathrm{s}}}{f_{\mathrm{H_2}}}
    =
    \exp\left(-\frac{\Delta \hat{G}^\circ}{RT}\right),
    \label{eqn:keqH2}
\end{equation}
where $f_{\mathrm{H_2}}=x_{\mathrm{H_2},\mathrm{g}}(\Gamma P/P^\circ)$ is the fugacity of gaseous hydrogen, $\Gamma$ is the fugacity coefficient, $a_{\rm H_2, s}$ is the activity of H$_2$ in the silicate melt, and $P^\circ$ is the standard-state pressure. 
In this formulation, the standard-state Gibbs free energy of dissolved \ce{H2} in the melt is computed from the combined fit anchored to the experimental constraints of \citet{hirschmann_solubility_2012} and the melt--gas chemical-potential equivalence from \citet{gilmore_coreenvelope_2026}, following \citet{werlen_effects_2026}.

The fixed-$k_{\rm D}$ treatment was adopted in \citet{schlichting_chemical_2022}, \citet{young_earth_2023}, \citet{werlen_atmospheric_2025}, and \citet{werlen_sub-neptunes_2025}. 
The fixed-$k_{\rm eq}$ treatment was adopted in \citet{werlen_effects_2026}. 
Given the current uncertainties, the choice of hydrogen-solubility prescription remains a model assumption.

Gibbs free energies for silicate and metal species are generally evaluated at a reference pressure of 1~bar. Their explicit pressure dependence is therefore neglected, as pressure-dependent thermodynamic data are generally not available over the relevant parameter range. Non-ideality in the gas and silicate phases is likewise not included. Introducing non-ideality in only one phase has been shown to produce internally inconsistent results. A more detailed discussion of pressure effects and non-ideality corrections in the gas and silicate phases is given in \citet{werlen_effects_2026}.

Non-ideal behavior is included for selected species in the metal phase. Activity coefficients for silicon and oxygen follow \citet{badro_core_2015} as implemented in \citet{Young+2023}, while carbon activity coefficients are adopted from \citet{fischer_carbon_2020} following \citet{werlen_atmospheric_2025}. When extending the reaction network, thermodynamically consistent Gibbs free energies must be provided for any additional reactions.

\section{\revisedtext{Comparison of different hydrogen solubility treatments}}
\label{app_comp}
\begin{figure}
    \centering
    \includegraphics[width=1.0\linewidth]{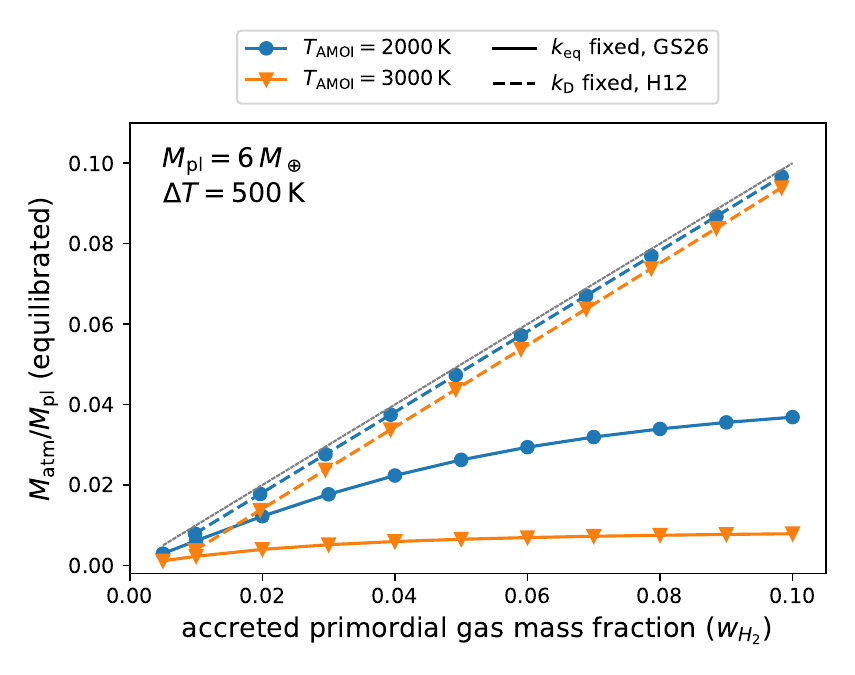}
    \caption{\revisedtext{Atmospheric mass fraction as a function of the accreted \ce{H2} mass fraction ($w_{\ce{H2}}$) for different AMOI temperatures and hydrogen-solubility prescriptions. Solid lines show the fixed-$k_{\rm eq}$ prescription, using the combined standard-state Gibbs free energy for dissolved \ce{H2} in the melt based on \citet{hirschmann_solubility_2012} and \citet{gilmore_coreenvelope_2026}. Dashed lines show the fixed-$k_{\rm D}$ prescription, using the pressure-independent reference-state Gibbs free energy from \citet{hirschmann_solubility_2012}. For all planets, $M_\mathrm{pl}=6\,M_\oplus$, $\Delta T=500\,$K, and $w_{\ce{H2O}}=0.0wt\%$.} }
    \label{fig:comp_Matm_H2_TAMOI_DeltaT}
\end{figure}

\revisedtext{
The treatment of hydrogen solubility in the silicate melt is a key model assumption in our work. 
We therefore compare the influence of the fixed-$k_{\rm D}$ prescription (used in e.g., \citet{SchlichtingYoung2022}) and the fixed-$k_{\rm eq}$ prescription used in this work on the atmospheric mass fraction, atmospheric metal mass fraction, and atmospheric C/O ratio in more detail. 
}

\begin{figure*}
        \centering
    \includegraphics[width=1.0\linewidth]{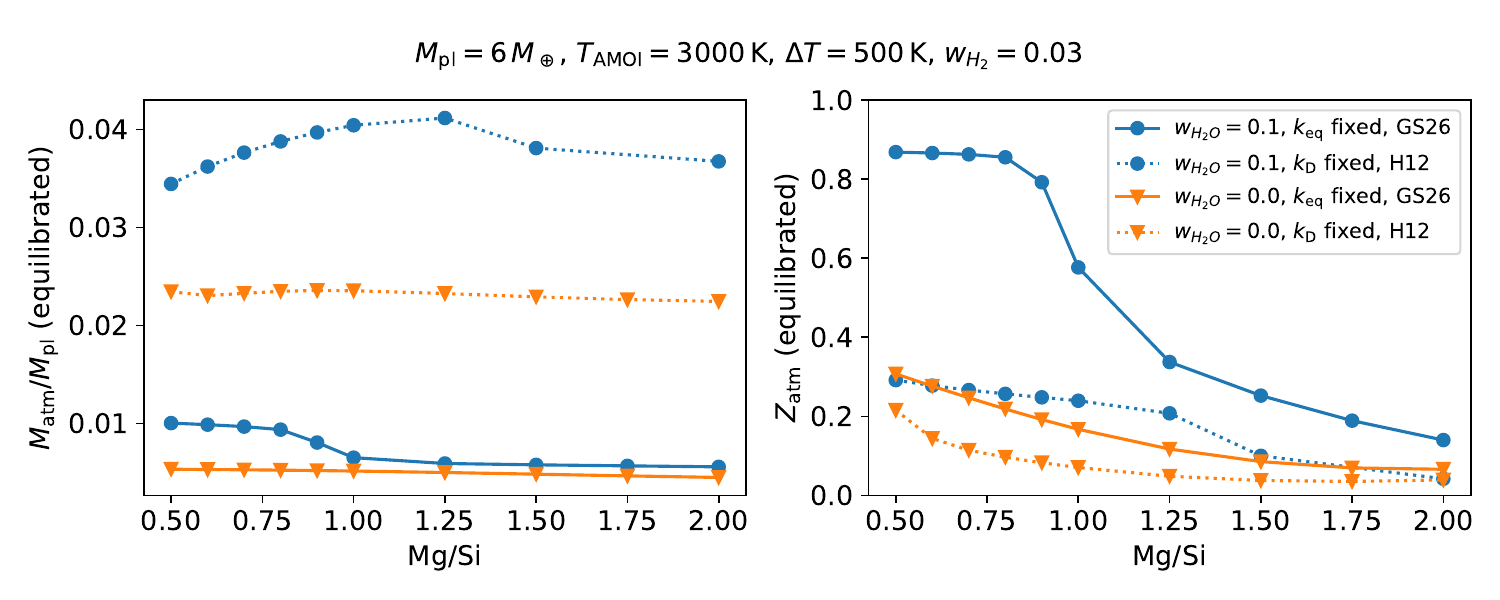}
    \caption{\revisedtext{Comparison of the atmosphere mass fraction (left plot) and metal mass fraction (right plot) as a function of molar Mg/Si for the two solubility treatments. Orange lines show planets with $w_{\mathrm{H}_2\mathrm{O}}=0.0wt\%$ and blue lines correspond to $w_{\mathrm{H}_2\mathrm{O}}=10wt\%$. The total mass is $6\,M_\oplus$,  $TAMOI=3000\,$K and $\Delta T = 500 K$ for all planets. Solid lines show the fixed-$k_{\rm eq}$ prescription, and dashed lines show the fixed-$k_{\rm D}$ prescription.}}
    \label{fig:comp_Matm_Z_MgSi_FeSi}
\end{figure*}

\revisedtext{
Figure~\ref{fig:comp_Matm_H2_TAMOI_DeltaT} illustrates the sensitivity of the atmospheric mass fraction to the hydrogen-solubility prescription.
The fixed-$k_{\rm D}$ prescription predicts that most of the hydrogen remains in the gas phase, resulting in atmospheric mass fractions which are similar to the total accreted \ce{H2} mass fraction. 
In contrast, the $k_{\rm eq}$ prescription produces atmospheric mass fractions that are significantly smaller or equal to the fixed-$k_{\rm D}$ prescription because it allows more efficient dissolution of \ce{H2} into the silicate phase at high pressure. 
In this case, the atmospheric mass fraction remains substantially smaller than the total accreted \ce{H2} mass fraction, indicating that most hydrogen is dissolved into the planetary interior rather than remaining in the atmosphere.
The phase in which the dissolved hydrogen resides in the interior depends on the adopted solubility prescription. For the fixed-$k_{\rm eq}$ prescription, most of the non-atmospheric hydrogen is stored in the silicate phase, primarily as dissolved \ce{H2} and \ce{H2O}. 
In contrast, for the fixed-$k_{\rm D}$ prescription, a larger fraction of the dissolved hydrogen partitions into the metallic phase.
}
\revisedtext{
For both solubility prescriptions, planets that accreted water in addition to primordial gas achieve higher atmospheric mass fractions (Figure~\ref{fig:comp_Matm_Z_MgSi_FeSi}). 
However, for the fixed-$k_{\rm D}$ prescription the atmospheric mass fraction of planets that accreted water is only weakly dependent on the rock composition. 
}
\revisedtext{
Figure~\ref{fig:comp_Matm_Z_MgSi_FeSi} further illustrates that the higher amount of hydrogen in the gas phase for the fixed-$k_{\rm D}$ prescription strongly reduces the atmospheric metall mass fraction compared to the fixed-$k_{\rm eq}$ prescription. 
The difference is more significant for planets with $w_{\ce{H2O}}=10wt\%$ and Mg/Si ratios below unity in the rock composition. 
For these planets, the atmospheric metal mass fractions is a factor three lower for the fixed fixed-$k_{\rm D}$ prescription. 
Nevertheless, the dependence of the atmospheric metal mass fraction on the Mg/Si ratio in the rocky building blocks is similar for both solubility prescriptions. 
For planets formed dry, the atmospheric metal mass decreases smoothly with increasing Mg/Si, while for planets that accreted water the atmospheric metal mass fraction sharply decreases once the Mg/Si is above unity. 
However, this decrease is more proununced for the fixed-$k_{\rm D}$ prescription.
}

\begin{figure*}
    \centering
    \includegraphics[width=1\linewidth]{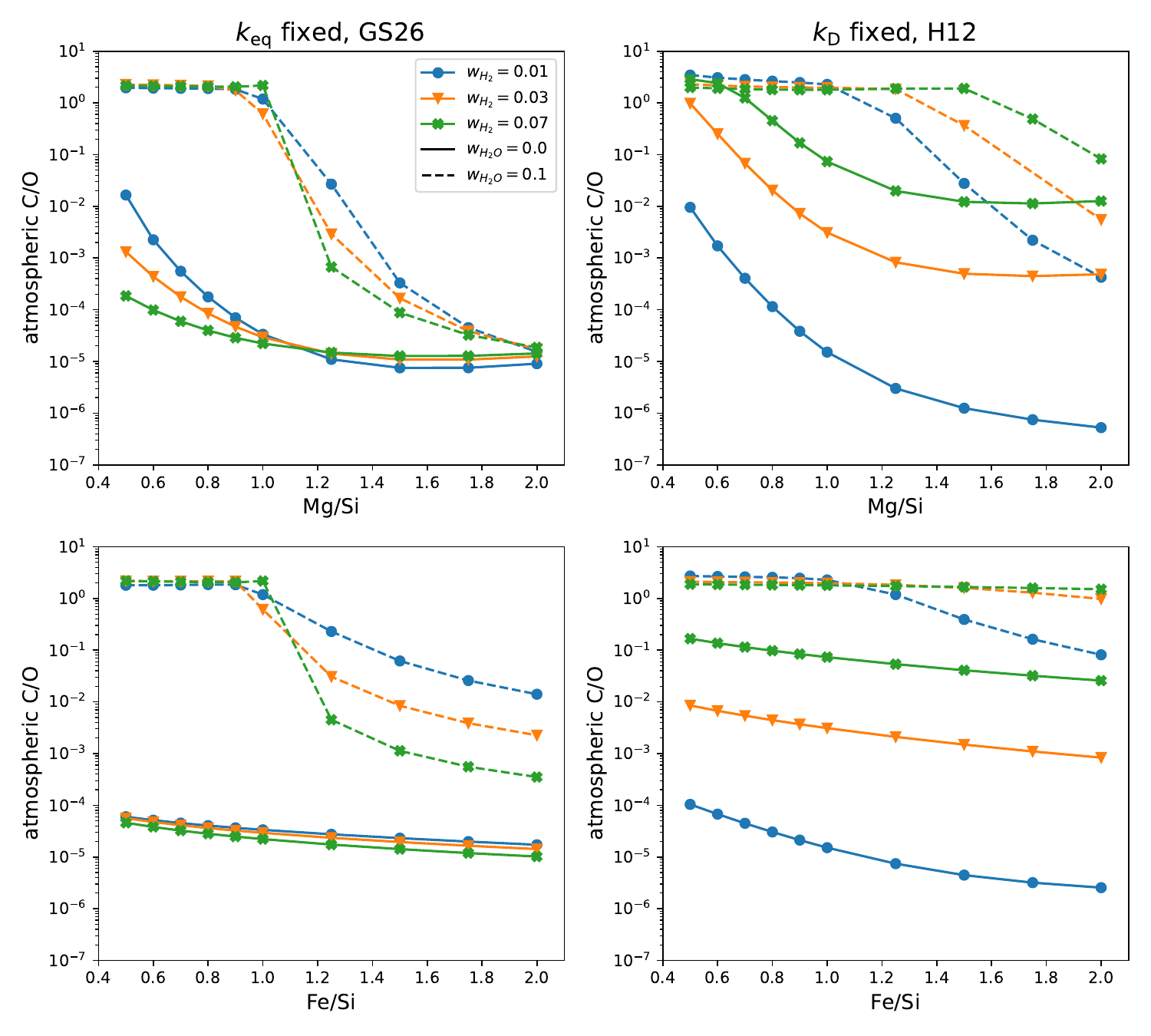}
    \caption{\revisedtext{Comparison of the atmospheric C/O ratio as a function of Mg/Si (top plot) and Fe/Si (bottom plot). The left column shows the fixed-$k_{\rm eq}$ prescription and the right column the fixed-$k_{\rm D}$ prescription. The colors and symbols show different mass fractions of H$_2$. Solid lines correspond to planets with $w_{\mathrm{H}_2\mathrm{O}}=0.0wt\%$ and dashed lines to planets with $w_{\mathrm{H}_2\mathrm{O}}=10wt\%$.}}
    \label{fig:comp_atmosphericCO}
\end{figure*}
\revisedtext{
Figure~\ref{fig:comp_atmosphericCO} shows the dependence of the atmospheric C/O ratio on the hydrogen solubility. 
Under the fixed-$k_{\rm eq}$ prescription adopted in this study, the majority of hydrogen partitions into the interior.
Consequently, the atmospheric C/O is relatively independent of $w_{\ce{H2}}$. 
In contrast, under the fixed-$k_{\rm D}$ prescription, the atmospheric C/O ratio increases with increasing $w_{\ce{H2}}$ as it leads to enhanced \ce{CH4} abundances in the atmosphere, see also \citet{werlen_atmospheric_2025}. 
This effect is strongest for planets with $w_{\ce{H2O}} = 0.0 wt\%$. 
Under the fixed-$k_{\rm D}$ prescription, the atmospheric C/O ratio does not depend on the Fe/Si ratio.
As a result, the distinct drop in the atmospheric metal mass fraction once Fe/Si reaches unity does not in appear in the atmospheric metal mass fraction. 
In contrast, the dependence of the atmospheric C/O ratio on the Mg/Si ratio is independent of the solubility prescription. 
}

\revisedtext{
This comparison highlights the importance of the adopted \ce{H2} solubility prescription on th atmospheric mass fraction, the atmospheric composition, metal mass fraction, and C/O ratio.
The results from equilibrium chemistry models in general should be interpreted within the framework of their employed hydrogen solubility treatment.
}

\bibliography{references}{}
\bibliographystyle{aasjournalv7}
\end{document}